\documentclass[twocolumn,superscriptaddress,prl,amsmath,amstex,amssymb,citeautoscript,longbibliography]{revtex4-1}
\pdfoutput=1
\usepackage{rotating}
\usepackage{natbib}
\usepackage[english]{babel}
\usepackage{letltxmacro}
\usepackage{latexsym}
\usepackage[utf8]{inputenc}
\LetLtxMacro{\ORIGselectlanguage}{\selectlanguage}
\makeatletter
\DeclareRobustCommand{\selectlanguage}[1]{%
  \@ifundefined{alias@\string#1}
    {\ORIGselectlanguage{#1}}
    {\begingroup\edef\x{\endgroup
       \noexpand\ORIGselectlanguage{\@nameuse{alias@#1}}}\x}%
}
\newcommand{\definelanguagealias}[2]{%
  \@namedef{alias@#1}{#2}%
}
\makeatother
\definelanguagealias{en}{english}
\definelanguagealias{English}{english}
\usepackage{graphicx}
\usepackage{amsmath}
\usepackage{amsfonts}
\usepackage{amsthm}   
\usepackage{amssymb}
\usepackage{bm}
\usepackage{color}
\usepackage[percent]{overpic}
\usepackage{soul} 
\usepackage{amssymb}
\usepackage{wasysym}
\usepackage{dsfont}
\usepackage{float}

\usepackage{hyperref}
\hypersetup{
    bookmarks=false,         
    unicode=false,          
    pdftoolbar=false,        
    pdfmenubar=true,        
    pdffitwindow=false,     
    pdfstartview={FitH},    
    pdftitle={},    
    pdfauthor={Authors},     
    pdfsubject={},   
    pdfcreator={},   
    pdfproducer={}, 
    pdfkeywords={quantum many-body scars} {eigenstate thermalization hypothesis} {non-equilibrium dynamics}, 
    pdfnewwindow=true,      
    colorlinks=true,       
    linkcolor=black,          
    citecolor=blue,        
    filecolor=magenta,      
    urlcolor=blue           
}

\newcommand{\be}{\begin{equation}}
\newcommand{\ee}{\end{equation}}
\newcommand{\bea}{\begin{eqnarray}}
\newcommand{\eea}{\end{eqnarray}}

\usepackage{graphicx}
\usepackage[colorinlistoftodos]{todonotes}
\usepackage{verbatim}

\newcommand{\tr}{\mathrm{tr}}

\usepackage{soul}
\usepackage[normalem]{ulem}

\newtheorem*{lemma*}{Lemma}

\begin{document}

\title{Systematic construction of scarred many-body dynamics in 1D lattice models}

\author{Kieran Bull}
\affiliation{School of Physics and Astronomy, University of Leeds, Leeds LS2 9JT, United Kingdom}
\author{Ivar Martin}
\affiliation{Materials Science Division, Argonne National Laboratory, Argonne, Illinois 60439, USA}
\author{Z. Papi\'c}
\affiliation{School of Physics and Astronomy, University of Leeds, Leeds LS2 9JT, United Kingdom}

\begin{abstract}
We introduce a family of non-integrable 1D lattice models that feature robust periodic revivals under a global quench from certain initial product states, thus generalizing the phenomenon of many-body scarring recently observed in Rydberg atom quantum simulators. Our construction is based on a systematic embedding of the single-site unitary dynamics into a kinetically-constrained many-body system.  We numerically demonstrate that this construction yields new families of models with robust wave-function revivals, and it includes kinetically-constrained quantum clock models as a special case. We show that scarring dynamics in these models can be decomposed into a period of nearly free clock precession and an interacting bottleneck, shedding light on their anomalously slow thermalization when quenched from special initial states.  
 \end{abstract}

\maketitle

{\sl Introduction.---}The understanding of ergodicity and thermalization in isolated quantum systems is an open problem in many-body physics, with important implications for a variety of experimental systems~\cite{Kinoshita06, Bloch15, Monroe16, Lukin16, Kaufman2016}. On the one hand, this problem has inspired important developments such as {\it Eigenstate Thermalization Hypothesis} (ETH)~\cite{DeutschETH,SrednickiETH,RigolNature}, which establishes a link between ergodicity and the properties of the system's eigenstates. On the other hand, strong violation of ergodicity can result in rich new physics, such as in  integrable systems~\cite{Sutherland}, Anderson insulators~\cite{Anderson58}, and many-body localized phases~\cite{Basko06,Serbyn13-1,Huse13}. In these cases, the emergence of many conservation laws prevents the system, initialized in a random state, from fully exploring all allowed configurations in the Hilbert space, causing a strong ergodicity breaking. 

A recent experiment on an interacting quantum simulator~\cite{Bernien2017} has reported a surprising observation of quantum dynamics that is suggestive of \emph{weak} ergodicity breaking. Utilizing large 1D chains of Rydberg atoms~\cite{Schauss2012,Labuhn2016,Bernien2017}, the experiment showed that quenching the system from a N\'eel initial state lead to persistent  revivals of local observables, while other initial states exhibited fast equilibration without any revivals. The stark sensitivity of the system's dynamics to the initial states, which were all effectively drawn from an ``infinite temperature" ensemble, appeared at odds with ``strong'' ETH~\cite{dAlessio2016, Gogolin2016,Mori2018}.

In Ref.~\onlinecite{Turner2017, wenwei18TDVPscar} the non-ergodic dynamics of a Rydberg atom chain was interpreted as a many-body generalization of the classic phenomenon of \emph{quantum scar}~\cite{Heller84}. For a quantum particle in a stadium billiard, scars represent an anomalous concentration of the particle's trajectory around (unstable) periodic orbits in the corresponding classical system, which has an impact on optical and transport properties~\cite{Sridhar1991, Marcus1992, Wilkinson1996}. By contrast, in the strongly interacting Rydberg atom chain initialized in the N\'eel state, quantum dynamics remains concentrated around a small subset of states in the many-body Hilbert space, thus it is effectively  ``semiclassical"~\cite{wenwei18TDVPscar}. While recent works~\cite{Khemani2018,Choi2018} have shown that revivals can be significantly enhanced by certain perturbations to the system, a general understanding of the conditions that allow scars to occur in a many-body quantum system is still lacking. 
 
The observation of periodic dynamics was linked to the existence of atypical eigenstates at evenly spaced energies throughout the spectrum of the many-body system~\cite{Turner2017,TurnerPRB,Iadecola2019}. Similar phenomenology was previously found in 
the non-integrable AKLT model~\cite{Bernevig2017,BernevigEnt}, where highly-excited eigenstates with low entanglement have been analytically constructed. A few of such exact eigenstates are now also available for the Rydberg atom chain model~\cite{lin2018exact}. In a related development, it was proposed that atypical eigenstates of one Hamiltonian can be ``embedded'' into the spectrum of another, ETH-violating, Hamiltonian~\cite{ShiraishiMori}. However, although the collection of models that feature atypical eigenstates is rapidly expanding~\cite{Calabrese16, Konik1, Konik2, IadecolaZnidaric, Dalmonte2019}, including recent examples of topological phases~\cite{NeupertScars} and fractons~\cite{FractonScars}, their relation to periodic dynamics remains largely unclear. 

In this paper we systematically construct interacting lattice models that exhibit periodic quantum revivals when quenched from a particular product state. 
The basic building block has a Hilbert space containing $N_c$ states (``colors") and a time-independent Hamiltonian that yields periodic unitary dynamics, $\mathcal{U}(t+T)=\mathcal{U}(t)$. The interacting models are defined by coupling these building blocks under a kinetic constraint.
Intriguingly, the dynamics in these models can be decomposed into periods of nearly free precession, in which the local degrees of freedom coherently cycle through the available states on a single site, followed by an interacting segment of dynamical evolution, reminiscent of a kicked quantum top~\cite{haake2001}. In all cases, the existence of atypical scarred eigenstates underpins the revivals. 
We show that our construction includes known models, such as chiral clock models~\cite{FendleyPara}, which are shown to support scars, and also gives a way of enhancing the revivals in spin-$s$ generalisations of the Rydberg chain~\cite{wenwei18TDVPscar}. Moreover, in selected cases with small values of $N_c$, we numerically explore general deformations of the models, verifying that our construction yields optimal models, i.e., with the highest amplitude of the wave function revivals.

{\sl PXP model.---}We start by briefly reviewing the model of a 1D Rydberg atom chain~\cite{Sun2008,LesanovskyDynamics,Olmos2012, Lesanovsky2012}. The system can be modelled as coupled two level systems (with
states $|0\rangle$, $|1\rangle$) described by an effective ``PXP'' Hamiltonian
\begin{eqnarray}\label{eq:pxp}
    H = \sum_j P_{j-1}^0 X_j P_{j+1}^0, \;\;\; P_j^0 \equiv |0_j\rangle\langle 0_j|,
\end{eqnarray}
where $X_j =|0_j \rangle\langle 1_j| + |1_j\rangle\langle 0_j|$ denotes the Pauli matrix.  
The model in Eq.~(\ref{eq:pxp}) describes a kinetically constrained paramagnet~\cite{Lesanovsky2018}: each atom can flip only if both its neighbors are in $\vert 0 \rangle$ state. 

The Hamiltonian in Eq.~(\ref{eq:pxp}) is interacting and non-integrable~\cite{Turner2017}, yet it exhibits unconventional thermalization. For example, the model has atypical (ETH-violating) eigenstates with low entanglement at high energy densities~\cite{TurnerPRB}. Moreover, when the system is quenched from the N\'eel initial state, $|\psi_0\rangle = |0101\ldots\rangle$, local observables such as domain wall density~\cite{Bernien2017} and even the many-body wave function fidelity, $F(t)=|\langle \psi_0 | \psi(t)\rangle|^2$, all revive with the same frequency~\cite{Turner2017, Iadecola2018, Dalmonte2019}. At the same time, quenches from other initial states, such as $|0000\ldots\rangle$, do not lead to observable revivals~\cite{Bernien2017}. The revival frequency from the N\'eel state is set by the energy separation between atypical eigenstates, as the same eigenstates also maximize the overlap with the N\'eel state~\cite{Turner2017}. Thus, the quench dynamics from the N\'eel state is largely restricted to a subset of many-body eigenstates, and can be viewed as precession of a large spin, which traces a periodic orbit that can be accurately captured by time-dependent variational principle (TDVP) on a manifold spanned by matrix product states with low bond dimension~\cite{wenwei18TDVPscar}.

{\sl Construction of scarred models.---} Consider now a system with a local basis $|0\rangle$, $|1\rangle$, ..., $|N_c-1\rangle$, and an arbitrary time independent Hamiltonian $h$ whose unitary dynamics is periodic, such that $\mathcal{U}^T \equiv \exp(-i h T) = \mathds{I}$ for arbitrary $T$ (not necessarily integer). The eigenvalues of $\mathcal{U}$ are then $\lambda_n = \exp(i 2\pi k_n/T)$, with the corresponding eigenvectors $\vert \psi_n \rangle$, where $k_n$ are arbitrary integers.  We can then obtain candidate Hamiltonians $h$ by choosing particular $\{ \lambda_n \}$ which guarantee a periodic $\mathcal{U}$ and taking its logarithm: 
\begin{eqnarray}\label{eq:c}
    h = i\sum_{n=0}^{N_c-1} \frac{2 \pi i}{T}k_n \; \vert \psi_n \rangle \langle \psi_n \vert.
\label{eq:clock_op}
\end{eqnarray}
To define a many-body lattice Hamiltonian, we take a tensor product of $h$ and impose the kinetic constraint that $h$ only acts on sites whose neighbors are in some unlocking state $\vert \phi \rangle$: 
\begin{eqnarray}
    H = \sum_{j=0}^{N-1} P_{j-1} h_j P_{j+1},\quad P_j \equiv \vert \phi_j \rangle \langle \phi_j \vert,
    \label{eq:pcpgeneral}
\end{eqnarray}
where $N$ is the number of lattice sites. The only other condition we place on $h$ is that the many-body system possesses a particle-hole symmetry $\rho$, which anticommutes with $H$,  $\{H,\rho\}=0$, leading to the symmetry $E \leftrightarrow -E$ of the energy spectrum. 
This is motivated by the fact that PXP model possesses such a symmetry, and its revivals are improved by the addition of perturbations which preserve this symmetry~\cite{Khemani2018,Choi2018}. Precise form of $\rho$ is unimportant here and can be found in \cite{SOM}.   We thus focus on cases where $\{k_n\}$ are symmetric around zero, resulting in $h$ being off diagonal and  compatible with $\rho$. A particularly illustrative example of this construction is when $\mathcal{U}$ is interpreted as the shift operator of a quantum clock~\cite{FendleyPara,Fendley2018}, as we explain next.

{\sl Scars in clock models.---}The scarred clock models are defined by choosing $T=N_c$, which gives
\begin{align}\label{eq:cdef}
\mathcal{U} = e^{-i C} = \sum_{j=0}^{N_c-1} |j+1\rangle\langle j|.
\end{align} 
In this case, $\lambda_n = \exp(2 \pi i k_n / N_c)$ and $ \vert \psi_n \rangle = \sum_{j=0}^{N_c-1} (1/\lambda_n^j) \vert j \rangle$. For odd $N_c$, $k_n$ takes the values $-\frac{N_c-1}{2},\ldots, 0, \ldots, \frac{N_c-1}{2}$. For $N_c$-even, we need to double the period, $T=2N_c$, in order to make $h$ off-diagonal in the $\vert j \rangle$ basis. This allows to choose $k=-\frac{N_c-1}{2},\ldots,-\frac{1}{2},\frac{1}{2},\ldots,\frac{N_c-1}{2}$, and Eq.~(\ref{eq:cdef}) continues to be valid for $N_c$-even after performing a gauge transformation, $|j\rangle\to e^{i\pi j/N_c}|j\rangle$.

\begin{figure*}[t]
    \centering
    \includegraphics[width=0.99\linewidth]{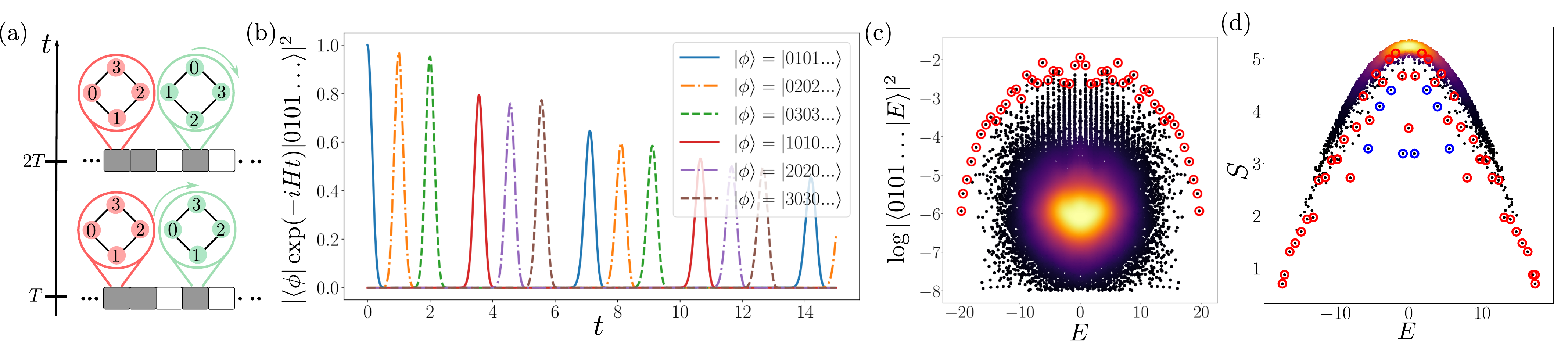}    
    \caption{(a)
A schematic of scarred clock models. A clock (green) can precess because both of its neighbors are in the unlocking state $|0\rangle$ (white), unlike the frozen clock (red).
(b) 
Dynamics of fidelity, $|\langle \phi |e^{-i t H}|1010\ldots\rangle|^2$, for $N_c=4$-color clock model in Eq.~(\ref{eq:pcp}). Different curves correspond to several choices of $|\phi\rangle$: the initial state $|1010\ldots\rangle$, the product state of shifted clocks $|n0n0\ldots\rangle$ ($n=2,3,\ldots,N_c-1$), or the overall translation of the initial state, $|0101\ldots\rangle$. (c) Overlap of all eigenstates of $N_c=4$-color clock model with the N\'eel state $|0101\ldots\rangle$. Each dot represents a (logarithm) of the overlap for an eigenstate $|E\rangle$ with energy $E$ shown on the $x$-axis. Color scale indicates the density of data points. Scarred states are marked by red circles.
(d) Entanglement entropy $S$ of all eigenstates of $N_c=4$-color clock model, plotted as a function of their energy $E$. Red circles indicate the matching scarred states from (c), while a few additional scar states, associated with the a ``defected $\mathds{Z}_4$" state, $\vert20002030103000\rangle$, are marked by blue circles. Plots (b), (c) are  for system size $N=16$, while (d) is for $N=14$. In all cases, we resolve translation and inversion symmetry, and plot both $[k=0, P=+]$ and $[k=\pi, P=-]$ sectors. 
    }
    \label{fig:clock_schem}
\end{figure*}

The inspiration behind Eq.~(\ref{eq:cdef}) is that local dynamics is a cyclic rotation around the basis of $N_c$ ``clock" states $\vert j \rangle$,  Fig.~\ref{fig:clock_schem}(a). The many-clock ``PCP"  Hamiltonian from Eq.~(\ref{eq:pcpgeneral}) then becomes 
\begin{eqnarray}\label{eq:pcp}
H_{\rm clock} = \sum_j P^0_{j-1} C_j P^0_{j+1},
\end{eqnarray}
with $C$ represents $h$ in Eq.~(\ref{eq:c}). Without loss of generality, we can choose the projector onto any of the clock basis states, e.g., $P^{0} =\vert 0\rangle \langle 0 \vert$. Thus, each site precesses around the clock of basis states if both its neighbors are in $\vert 0 \rangle$ state, otherwise the site remains frozen, Fig~\ref{fig:clock_schem}(a). 

We have studied the PCP model in Eq.~({\ref{eq:pcp}) using exact diagonalization with periodic boundary conditions. For any $N_c\leq 12$ accessible to us numerically, we find long-lived oscillatory dynamics when the system is quenched from any N\'eel-like state, $\vert 0101... \rangle$, $\vert 0202... \rangle$, etc. Fig.~\ref{fig:clock_schem}(b) summarizes the result for $N_c=4$. The dynamics proceeds in two steps. First,  each unfrozen clock nearly freely cycles through its states, $|1\rangle \to |2\rangle \to \ldots |N_c-1\rangle$. After this coherent process is complete, the many-clock state shifts, $|N_c-1,0,N_c-1,0\ldots \rangle \to |0101\ldots \rangle$. In this second step, interactions kick in and some fidelity is lost  to thermalization. This cycle of free precession followed by a short interacting segment is reminiscent of a kicked quantum top. We now see that the PXP model (which is equivalent to $N_c=2$ clock) is special in that it lacks free-precession dynamics. On the other hand, similar to the PXP case, in scarred clock models coherence also remains protected to a large degree during the interacting part of the process, allowing the  wave function to keep returning to the initial state at later times. 

In order to visualize the dynamics, in Fig.~\ref{fig:clock_schem}(b) we plot the fidelity $|\langle \phi |\exp(-i t H)|1010\ldots\rangle|^2$ w.r.t. several product states $|\phi\rangle$  corresponding to either the initial state, the internal shift of each clock, or to the overall translation of the initial state.  The duration of individual clock ticks (e.g., $|1010\ldots \rangle \to |2020\ldots\rangle$) matches that of the \emph{unconstrained} clock model. Following the convention that $C$ is rescaled such that nearest neighbor hoppings have magnitude one, the frequency of the putative free precession is found to be $\approx 0.902$ (in units $\hbar=1$) while the frequency of the single site precession (in the absence of a constraint) is $\approx 0.900$.  We note that time evolution of local observables is consistent with the presented picture of the underlying dynamics~\cite{SOM}.

In Fig.~\ref{fig:clock_schem}(c) we show the overlap of all eigenstates with the N\'eel state $|0101\ldots\rangle$, while Fig.~\ref{fig:clock_schem}(d) shows the bipartite entanglement entropy $S=-\tr\rho_A \ln \rho_A$, where $\rho_A$ is the reduced density matrix of one half of the chain. The scar states are easily identifiable as a band of special eigenstates (circled in red) that extend throughout the spectrum. Total number of special states is $(N_c-1)N+1$.
Similar to the PXP model, the special eigenstates are distinguished by their high overlap with the N\'eel state, or alternatively as ones with atypically low entanglement. We note that some of the eigenstates with smallest entanglement belong to a different band of scarred states associated with  a ``defected $\mathds{Z}_4$'' state $\vert 20002030103000\rangle$ (blue circles in Fig.~\ref{fig:clock_schem}(d)).  Apart from these special states, we also observe tower structures in the spectrum, which reflect the clustering of neighboring eigenstates around the energies of the scarred eigenstates. Deep in the bulk of the spectrum, the density of states [indicated by color scheme in Fig.~\ref{fig:clock_schem}(c)] appears to be uniform, as expected from the ETH. Indeed, at $N=14$ we find a mean level spacing ratio~\cite{OganesyanHuse} of $\langle r\rangle =0.5218$, consistent with Wigner-Dyson statistics and the ETH. We have confirmed that the frequency of the revival to the initial state matches the energy separation between special eigenstates in Fig.~\ref{fig:clock_schem}(c).

    {\sl Relation to spin-$s$ and chiral clock models.---}In Ref.~\cite{wenwei18TDVPscar} the TDVP approach was generalized to spin-$s$ PXP models with the kinetic constraint $P^0$. Periodic revivals were numerically demonstrated for $s=1, 2$. Both spin-$s$ PXP model and $N_c=2s+1$ colored PCP clock models are obtained from our construction in Eq.~(\ref{eq:pcpgeneral}) by taking $k=-s,...,s$. Thus by performing a basis rotation, the clock Hamiltonian can be expresed in the spin basis, $H_{\rm clock} = \sum_j P_{j-1}' X_j P_{j+1}'$, where $P'$ is a deformation of the projector $P^0$ in Eq.~(\ref{eq:pcp})~\cite{SOM}. We have numerically found that the number of scarred states remains the same for PXP models expressed  in terms of either the spin $P^0$ or $P'$; however, the amplitude of the revivals is always higher when using $P'$ instead of spin $P^0$ for $N_c$-odd~\cite{SOM}. Thus, our construction shows how to improve the revivals in the standard PXP models. In addition, mapping to the clock representation allows to clearly delineate nearly-free precession from the interacting part of the dynamics, which is not transparent in the spin representation.  

Furthermore, our construction includes models for which $C$ is not related to spin matrices via a change of basis. One family of models for even $N_c$ is obtained by  choosing $k=-\frac{N_c}{2},\ldots,-1,1,\ldots,\frac{N_c}{2}$, with the projector $P^0$ as above. For $N_c=4$, the resulting $C$ is equivalent to the 4-color Chiral Clock Model (CCM) at the fixed point in the disordered phase~\cite{Baxter1989, FendleyPara, SOM}. This model exhibits two types of oscillatory behaviour. Quenches from $\vert 0202...\rangle$ result in decaying fidelity revivals. Quenches from $\vert 1010...\rangle$, $\vert 3030...\rangle$ essentially freeze out the sublattice containing $0$'s, such that the system oscillates like a nearly free paramagnet~\cite{SOM}.
 
{\sl General phase diagram of scarred models.---}We now perform an extensive  search for scarred models with the fixed kinetic constraint $P^0$. By varying elements of $C$, we scan all models of the form Eq.~(\ref{eq:pcp}). We map out the phase diagram of these models based on the quality of scars, i.e., the first revival maximum of the fidelity from the N\'eel-like states. We restrict the matrix $C$ to be purely imaginary and off diagonal, as this leads to a symmetric spetrum~\cite{SOM}, thus preserving the desired particle-hole symmetry.

\begin{figure}[t]
    \centering
    \includegraphics[width=\linewidth]{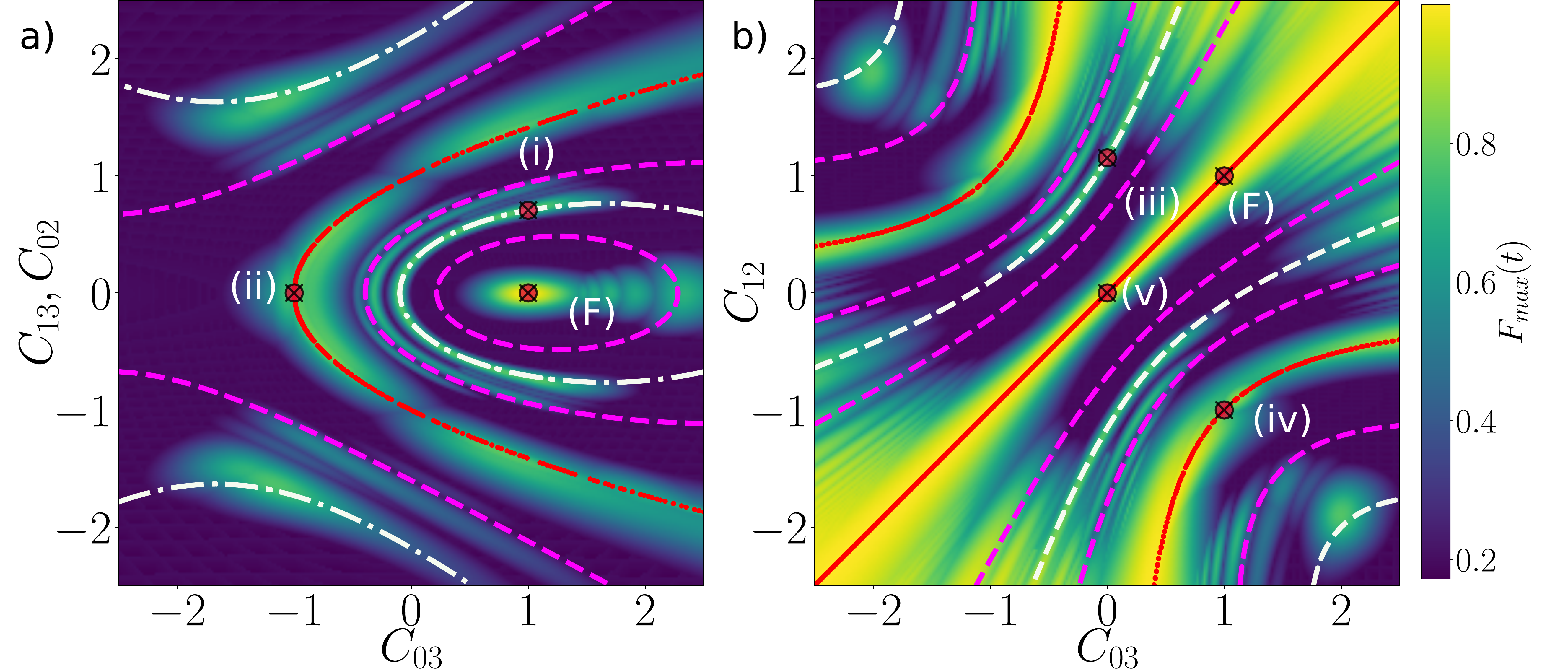}
    \caption{The phase diagram of scarred models with $N_c=4$ and projector $P^0$. Shown in (a), (b) are two slices of the phase diagram obtained by varying the matrix elements of $C$, indicated on the axes. Color scale represents the maximum of the first fidelity revival for quenches from any of the states $\vert0101\ldots\rangle$, $\vert0202\ldots\rangle$, $|0303\ldots\rangle$. Results are for system size $N=10$. Labels on the diagrams refer to special limiting cases defined in the text. Scarred models can be accurately predicted based on the commensurability of the eigenvalue spectrum of $C$, as denoted by lines and explained in the text. 
    }
    \label{fig:4colour_phase}
\end{figure}
Consider the $N_c=4$ case. Allowed distortions involve varying 5 matrix elements in $C$, so we take slices  where only two parameters are simultaneously varied. We consider two cases, (a) vary the next-nearest-neighbor hoppings $C_{02} = C_{13} = \alpha i$, while also varying $C_{03} = -\beta i$, or (b) switch off next-nearest-neighbor hoppings, while varying $C_{12} = -\alpha i $ and $C_{03} = -\beta i $. The corresponding phase diagrams are shown in Fig.~\ref{fig:4colour_phase}. On these diagrams there are several limiting cases at special values of $(\beta,\alpha)$. For variation (a), we have: (i) $(1,1/\sqrt{2})$ is $N_c = 4$ clock; (ii) $(-1,0)$ is $N_c=4$ CCM model. For variation (b): (iii) $(0,2/\sqrt{3})$ is spin-$\frac{3}{2}$ PXP; (iv) $(1,-1)$ is also $N_c=4$ CCM; (v) at $(0,0)$, we have $C=i\sum_{j=0,2} |j\rangle\langle j+1| - \mathrm{h.c.}$, which (with $P^0$) can be viewed as the sum of a spin-$\frac{1}{2}$ PXP and a free $s=\frac{1}{2}$ paramagnet. Points marked F correspond to decoupled free paramagnets. 

The maximum fidelity at first revival for $N_c$-even is generally comparable between clock and spin-$s$ PXP models. For example, for $N_c=4$ in Fig.~\ref{fig:4colour_phase},  $F_{max}\approx 0.761$ (clock) and $F_{max} \approx  0.783$ for spin-$\frac{3}{2}$ PXP. For $N_c=6$ and $N=8$, we obtain $F_{max} \approx 0.813$ (spin) and  $F_{max}\approx 0.802$ (clock), while for $N_c=8$, $N=8$ we find $F_{max} \approx 0.793$ (spin) and $F_{max} \approx 0.806$ (clock). On the other hand, for $N_c$-odd, we find a considerable improvement in the fidelity of a clock compared to the spin-$s$ PXP model. For example, for $N_c=3$, the maximum fidelity of the clock model is $F_{max} \approx  0.724$ versus $F_{max} \approx 0.653$ for spin-$1$; for $N_c = 5$, $N = 10$, the improvement is even bigger, $F_{max} \approx  0.563$ vs. $F_{max} \approx  0.766$ (clock)~\cite{SOM}. Thus, our construction for odd $N_c$ gives a way to improve the revivals over corresponding $s=(N_c-1)/2$ spin-$s$ PXP models.

Since the phase diagram in Fig.~\ref{fig:4colour_phase} is evidently quite rich, we look for a simple guiding principle that predicts the most robust scarring models. The commensurability of the eigenvalue spectrum of $C$ provides such a criterion, see lines and dots in Fig.~\ref{fig:4colour_phase}. White lines mark the models for which  $C$ has equidistant energy levels, i.e., its eigenvalues are $E_n = k \epsilon$, $k \in \mathds{Z}$. Our $N_c=4$ clock model lies on one of these lines, as shown in Fig.~\ref{fig:4colour_phase}(a). We can consider further commensurability conditions where the energy spacings of $C$ are in simple ratios, e.g., 1:2 (pink lines). Finally, red points mark the areas where $C$ contains one pair of degenerate eigenvalues (with trivially commensurate spacing in our case). One of these points hosts the $N_c=4$  CCM at its fixed point in the disordered phase. Another one, along the diagonal in Fig.~\ref{fig:4colour_phase}(b), hosts a combination of the free paramagnet and spin-$\frac{1}{2}$ PXP model. We note, however, that our simple criterion based on the non-interacting spectrum of $C$ only serves as a rough indicator of scarring models. The precise parameter values where such models are realized are determined by the non-trivial interplay between this condition and the kinetic constraint, i.e., $P^0$.

{\sl Conclusion.---}We have presented a systematic construction of non-integrable PCP models exhibiting many-body revivals and quantum scars. The construction is based on embedding local unitary precession, $U^T=e^{-iCT}=\mathbb{I}$, into an interacting quantum system.  The obtained models are expressed in terms of kinetic constraints which are present  in  quantum simulators in the Rydberg blockade regime~\cite{Bernien2017,Dalmonte2019,Keesling2019}. Kinetic constraints of this kind also emerge naturally in lattice gauge theories, which have recently been realized in periodically driven optical lattices~\cite{optical_lattices_kinetic}.
The strongest reviving models are predicted by considering the commensurability of $C$'s eigenvalues. For odd $N_c$ and equidistant eigenvalues for $C$, the obtained models revive better than the corresponding spin $s=(N_c-1)/2$ PXP model. Rotating $C\rightarrow X$, $P\rightarrow P^{'}$, our construction thus provides a prescription for improving PXP revivals. If we do not restrict to equidistant eigenvalues of $C$, our construction yields further families of scarred models not related to PXP by rotation. Further, clock models provide a simple physical picture of the underlying dynamics -- a period of nearly free precession followed by an interacting bottleneck. This ``effective drive" is reminiscent of kicked systems, where mixed phase space dynamics (both recurrent and thermalizing behaviour), can emerge due to the presence of a continuous spectrum in the Floquet operator~\cite{sc_spectrum_rotator}.
Taking the same constraint $U^T=\mathbb{I}$, one can also engineer time-translation symmetry breaking in driven systems~\cite{ttsb_eng_spin_chain,time_crystal_prethermal}. These observations suggest a deeper connection between oscillatory scarred models and time crystals, complementing recent description of scarred PXP states as $\pi$ magnon condensates which possess long range order in both space and time~\cite{Iadecola2019}. 

\begin{acknowledgements}
{\it Acknowledgements.---}We thank Paul Fendley for useful comments. K.B. and Z.P. acknowledge support by EPSRC grants EP/P009409/1 and EP/R020612/1. Statement of compliance with EPSRC policy framework on research data: This publication is theoretical work that does not require supporting research data. This research was supported in part by the National Science Foundation under Grant No. NSF PHY-1748958. Work at Argonne National Laboratory was supported by the
Department of Energy, Office of Science, Materials Science and Engineering Division.
\end{acknowledgements}
\bibliography{refs}

\clearpage 
\pagebreak

\onecolumngrid
\begin{center}
\textbf{\large Supplemental Online Material for ``Systematic construction of scarred many-body dynamics in 1D lattice models'' }
\end{center}

\vspace{0.1cm}
{\small In this supplementary material, we discuss in detail the relation between the clock models introduced in the main text, the spin-$s$ PXP models, and chiral clock models (CCMs). We complement the results for the fidelity dynamics in the main text with a study of dynamics of local observables following the quench. We also give the explicit form of discrete symmetries present in the clock models. Further we present a finite-size scaling analysis of fidelity revivals seen in clock models, and discuss the effect of boundary conditions. Moreover, we show that scarred eigenstates in clock models can be captured using a Krylov-type approximation scheme. Finally, we present two families of perturbations which improve revivals substantially in the clock models, while also considering the effect of these perturbations on the level statistics.
}
\setcounter{equation}{0}
\setcounter{figure}{0}
\setcounter{table}{0}
\setcounter{page}{1}
\setcounter{section}{0}
\makeatletter
\renewcommand{\theequation}{S\arabic{equation}}
\renewcommand{\thefigure}{S\arabic{figure}}
\renewcommand{\thesection}{S\Roman{section}}
\renewcommand{\thepage}{S\arabic{page}}
\vspace{0.5cm}
\twocolumngrid

\section{Revivals of local observables}

In the main text we  analyzed the dynamics in the clock models from the perspective of fidelity revivals, i.e., the many-body wave function returning to a particular product state after time $t$. By inspecting different product states, we have concluded that the dynamics proceeds in two steps: (i) precession of every clock between the available basis states; (ii) transition to a product state that is related to the initial state by an overall translation of the system by one lattice site, e.g., $|1010\ldots\rangle \to |0101\ldots\rangle$.  While the process (i) is nearly free (and hence coherent), remarkably even the process (ii) results in a relatively small loss of fidelity. 

Revival of a many-body wave function, of course, is formidable to measure experimentally, and here we demonstrate that the same picture of the dynamics can be inferred from studying revivals in local observables, $Q$. For example, measuring a color on a given site, $Q \equiv P^n = |n\rangle\langle n|$ will show the same pattern of dynamics, see Fig.~\ref{fig:orderparameter}. Consider the two sublattices, corresponding to even and odd sites. When the system is quenched from the N\'eel state $\vert 0101...\rangle$, even sites initially cycle through the basis set $1 \rightarrow 2 \rightarrow 3 ...$ while odd sites are locked at $0$ by the constraint. These sublattices exchange and the process repeats after a full precession $1\rightarrow0$ has been completed.
\begin{figure}[htb]
\includegraphics[width=\linewidth]{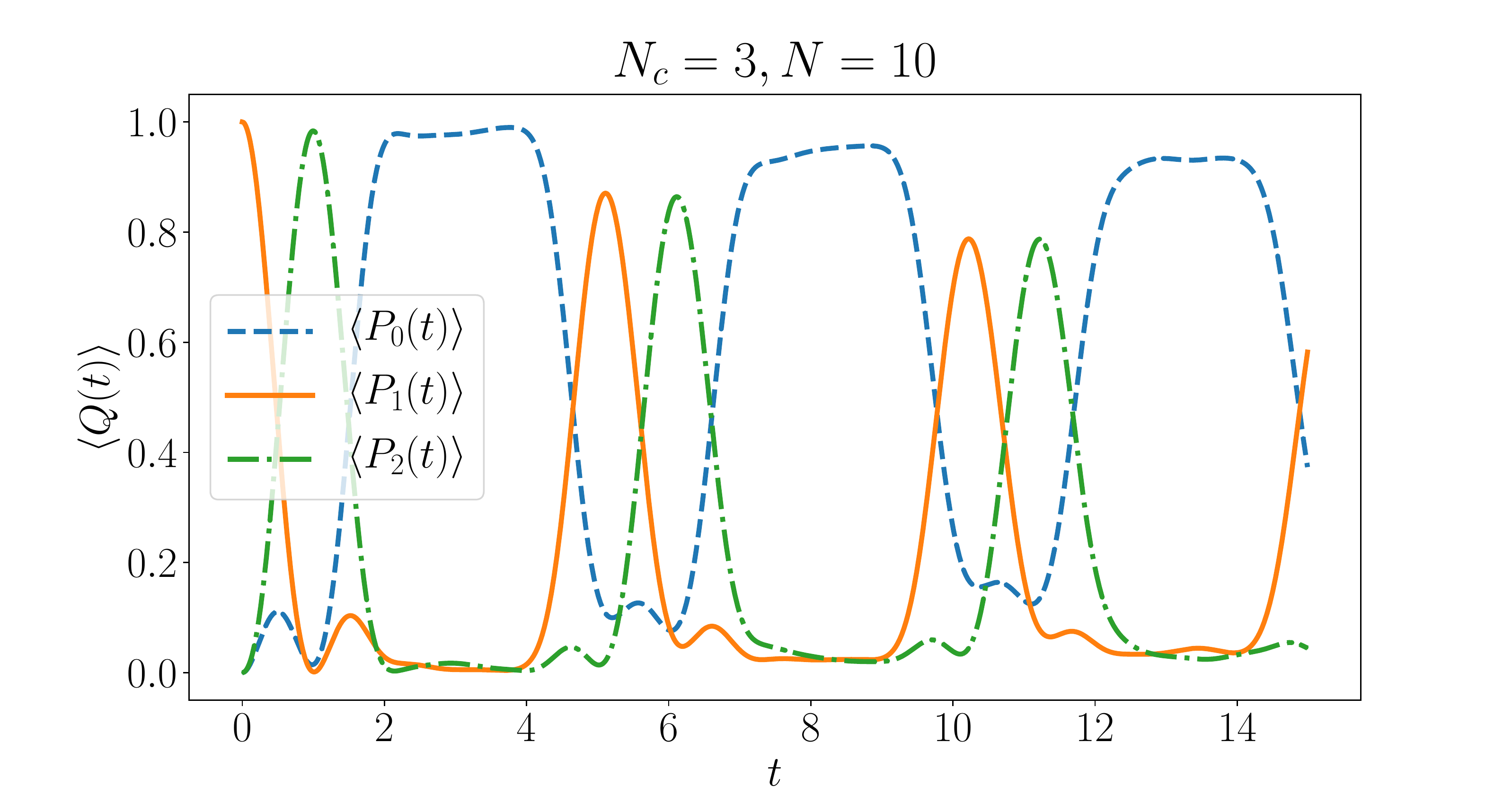}
\includegraphics[width=\linewidth]{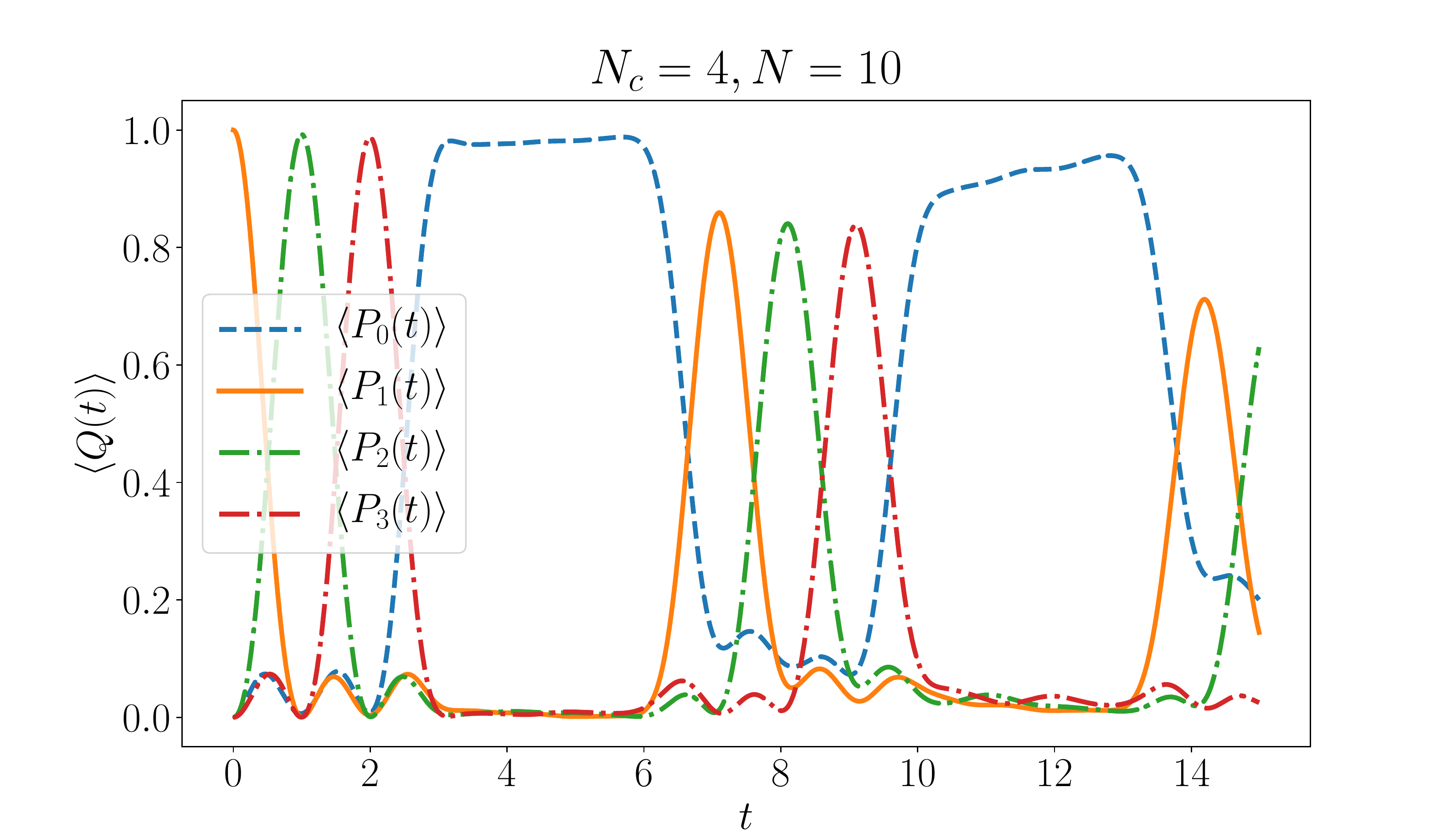}
\caption{Time evolution of local observables when the PCP model is quenched from the N\'eel state $\vert 0101...\rangle$. Plots show the time evolution of the average color on a given site, $\langle P^n(t)\rangle$, as a function of time. Top plot is for $N_c=3$ and bottom plot is $N_c=4$, with system size $N=10$. A period of free precession can be observed where the sublattice containing $1$ cycles through the clock basis, followed by a period of time where the color on the site is locked to $0$ while the other sublattice cyles through the clock basis.}
\label{fig:orderparameter}
\end{figure}

\section{Discreet Symmetries in Clock Models}

In the main text we noted our construction of the clock models yielded systems possessing a desired (anti-unitary) particle-hole symmetry. This was motivated by the fact that the original PXP model possesses a particle-hole symmetry $\rho = \prod_i Z_i$, which anticommutes with H, $\{H,\rho\}=0$. Our clock model actually possesses several discrete unitary and anti-unitary symmetries, originating from charge conjugation, parity and time reversal.
We define single-site charge conjugation as the following $N_c \times N_c$ operator~\cite{Whitsitt}:
\begin{align}
    \mathcal{C}_i = \begin{pmatrix} 1 & 0 & 0 & ... & 0 \\
        0 & 0 & ... & \pm1 & 0 \\
        0  & ... & \pm1 & 0 & 0 \\
        ... \\
        0 & \pm 1 & 0 & ... & 0
    \end{pmatrix},
\end{align}
where $+$ corresponds to odd $N_c$ while $-$ corresponds to even $N_c$. Further, we define parity and time reversal operators in the conventional way:
\begin{align}
    I:& \quad j \rightarrow L-j-1, \\
    T:& \quad \hat{K},
\end{align}
where $\hat{K}$ corresponds to complex conjugation. It then follows that the PCP models possess the following discreet symmetries:
\begin{align}
    [H,I] &= 0, \\
    [H,\mathcal{C}T] &= 0, \\
    \{H,\mathcal{C}\} &= 0, \\
    \{H,\mathcal{C}I\} &= 0,
\end{align}
where $\mathcal{C}=\prod_i\mathcal{C}_i$.

\section{Relation between clock and spin-$s$ PXP  models}

In the main text we pointed out that spin-$s$ PXP models are included in our construction as a special case.  From the definition of $\mathcal{U}$ governing the free precession of our clock models, which we reproduce here,
\begin{align}\label{eq:cdef}
\mathcal{U} = e^{-i C} = \sum_{j=0}^{N_c-1} |j+1\rangle\langle j|,
\end{align} 
 the eigenvalues are found to be
 $\lambda_n = \exp(2 \pi i k_n / N_c)$, with $k_n \in \{ -(N_c-1)/2,\ldots, 0, \ldots, (N_c-1)/2\}$ for odd $N_c$, while $k_n \in \{-(N_c-1)/2,\ldots,-1/2,1/2,\ldots,(N_c-1)/2\}$ for even $N_c$. This then implies the eigenvalues of $C\propto i \ln(\mathcal{U})$ are proportional to $k_n$, which we immediately recognize as angular momenta quantum numbers for spin $2s+1=N_c$. This tells us that operators $C$ and $X$ are equivalent by a change of basis.

 The distinction between the clock and spin models thus lies entirely in the kinetic constraint, i.e., the choice of the projector $P = \vert \phi \rangle \langle \phi \rangle$. For example, at $N_c=3$, we find:
 \begin{align}
     H_{clock} &= \sum_n P^0_{n-1} C_n P^0_{n+1},\;\;\; P^0 = \begin{pmatrix} 1 & 0 & 0 \\ 0 & 0 & 0 \\ 0 & 0 & 0 \end{pmatrix}\\
                  &= \sum_n P^{'}_{n-1} X_n P^{'}_{n+1},\;\;\;
      P^{'} = \begin{pmatrix} 1/6 & 1/3 & -1/6 \\ 1/3 & 2/3 & -1/3 \\ -1/6 & -1/3 & 1/6 \end{pmatrix}.
 \end{align}
For $N_c=3$, it is straightforward to visualize the entire phase diagram with the fixed projector $P^0$. There are two couplings to vary, as by a rescaling of $C$ we are free to fix the value of one of the three couplings. The resulting phase diagram for  $N_c=3$ is shown in Fig.~\ref{fig:3colour_phase} as a function of $C_{02}$ and $C_{12}$ matrix elements.
\begin{figure}[h]
    \centering
    \includegraphics[width=0.35\textwidth]{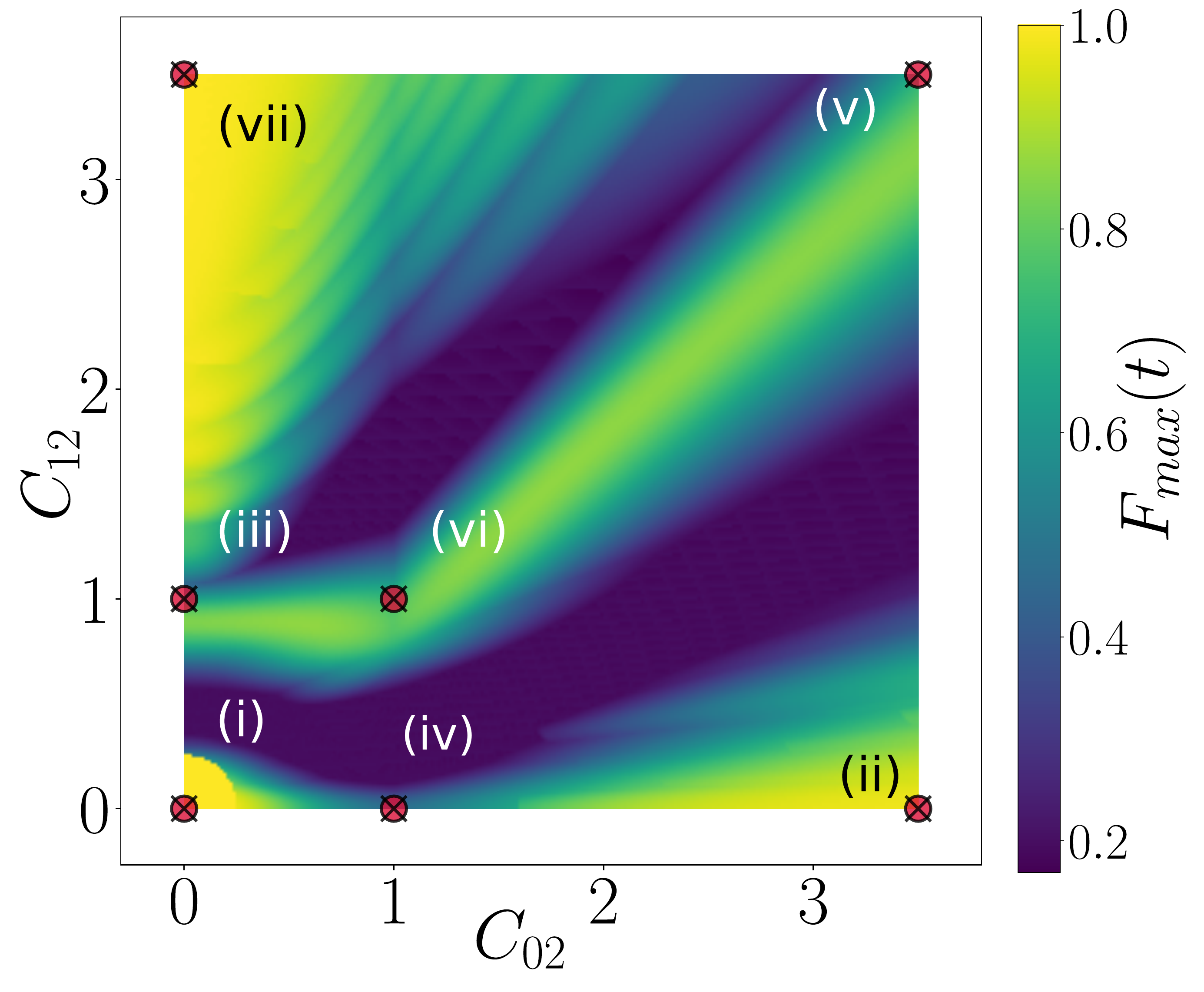}
    \caption{Phase diagram of the $N_c=3$ models obtained by varying the two matrix elements, $C_{02}$ and $C_{12}$. Color scale indicates the average maximum of the first fidelity revival quenched from the N\'eel states $|0101\ldots\rangle$ and $|0202\ldots\rangle$. All regions associated with strong scarring can be identified with either the spin or clock models.
      }
    \label{fig:3colour_phase}
\end{figure}
All the regions in the diagram that display  strong scarring can be interpreted as deformations of one of our previously discussed models. Specifically, in Fig.~\ref{fig:3colour_phase} we identify the representative points as: (i) and ii) are spin-$\frac{1}{2}$ PXP, (iii) is spin$-1$ PXP, with $P=\vert S_z=-1 \rangle \langle S_z=-1 \vert$, (iv) spin-$1$ PXP, with $P=\vert S_z=0 \rangle \langle S_z=0 \vert$, (v) spin-$1$ PXP, with $P=\vert S_z=1 \rangle \langle S_z=1 \vert$, (vi) $N_c=3$-color clock model, (vii) free paramagnet. From this we conclude that for the on-site Hilbert space dimension $N_c=3$, the only scarred models are obtained by various ways one can constrain a free paramagnet. As mentioned in the main text, the maximum fidelity of the clock model, $0.724$, is greater than the spin-$1$ model, $0.653$. This improvement in the revival fidelity is seen in all $N_c$-odd models. For example, at $N_c=5$, $N=10$, spin: $0.563$, clock: $0.766$. Thus, by constraining the free paramagnet with $P^{'}$ as opposed to $P^0$ leads to better revivals for $N_c$ odd. Finally, although still expressable as spin models, the clock basis provides a much simpler interpretation of the dynamics, clearly showing a period of free precession followed by an interacting segment.

\begin{figure}[htb]
\includegraphics[width=0.48\linewidth]{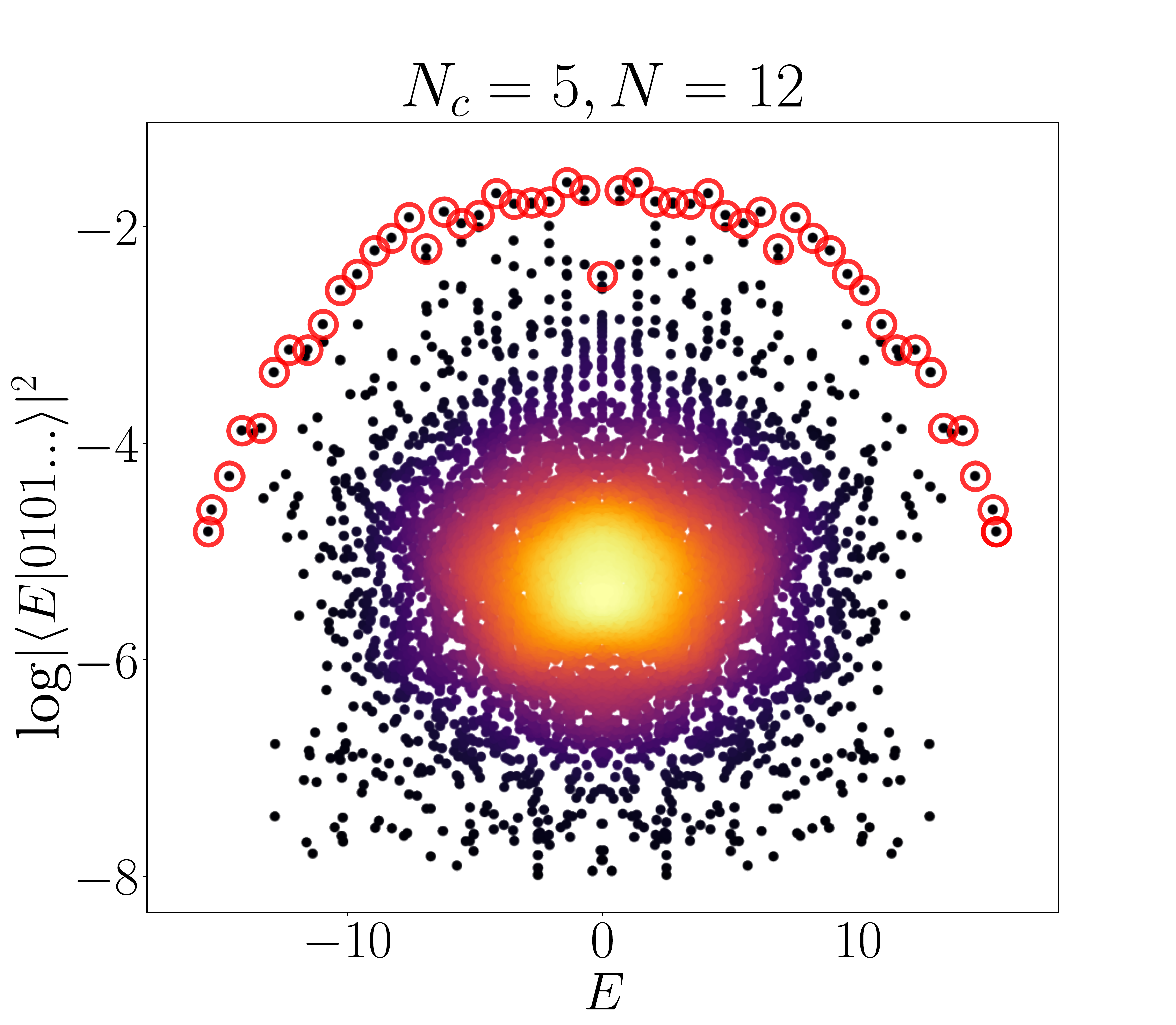}
\includegraphics[width=0.48\linewidth]{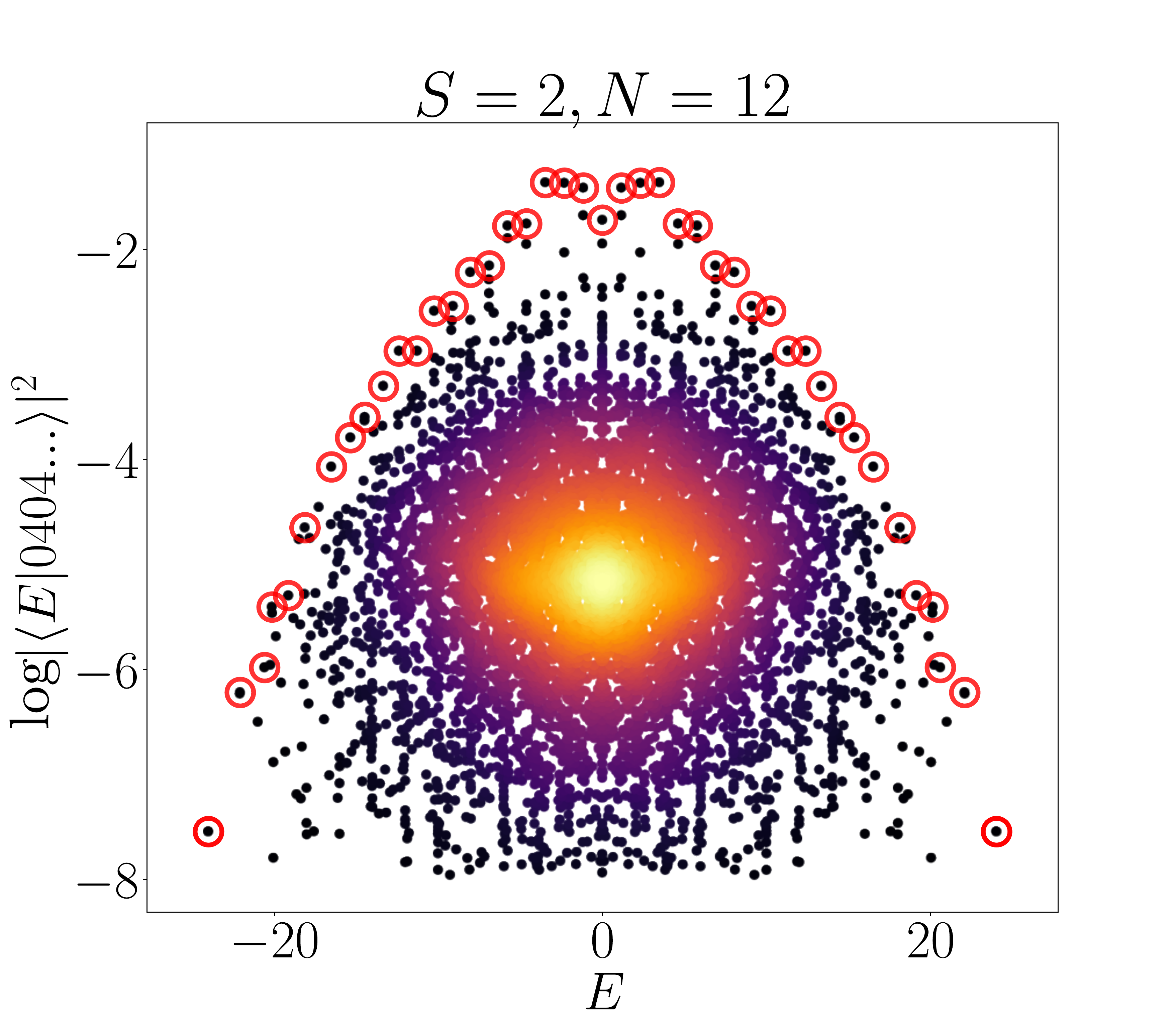}
\includegraphics[width=0.48\linewidth]{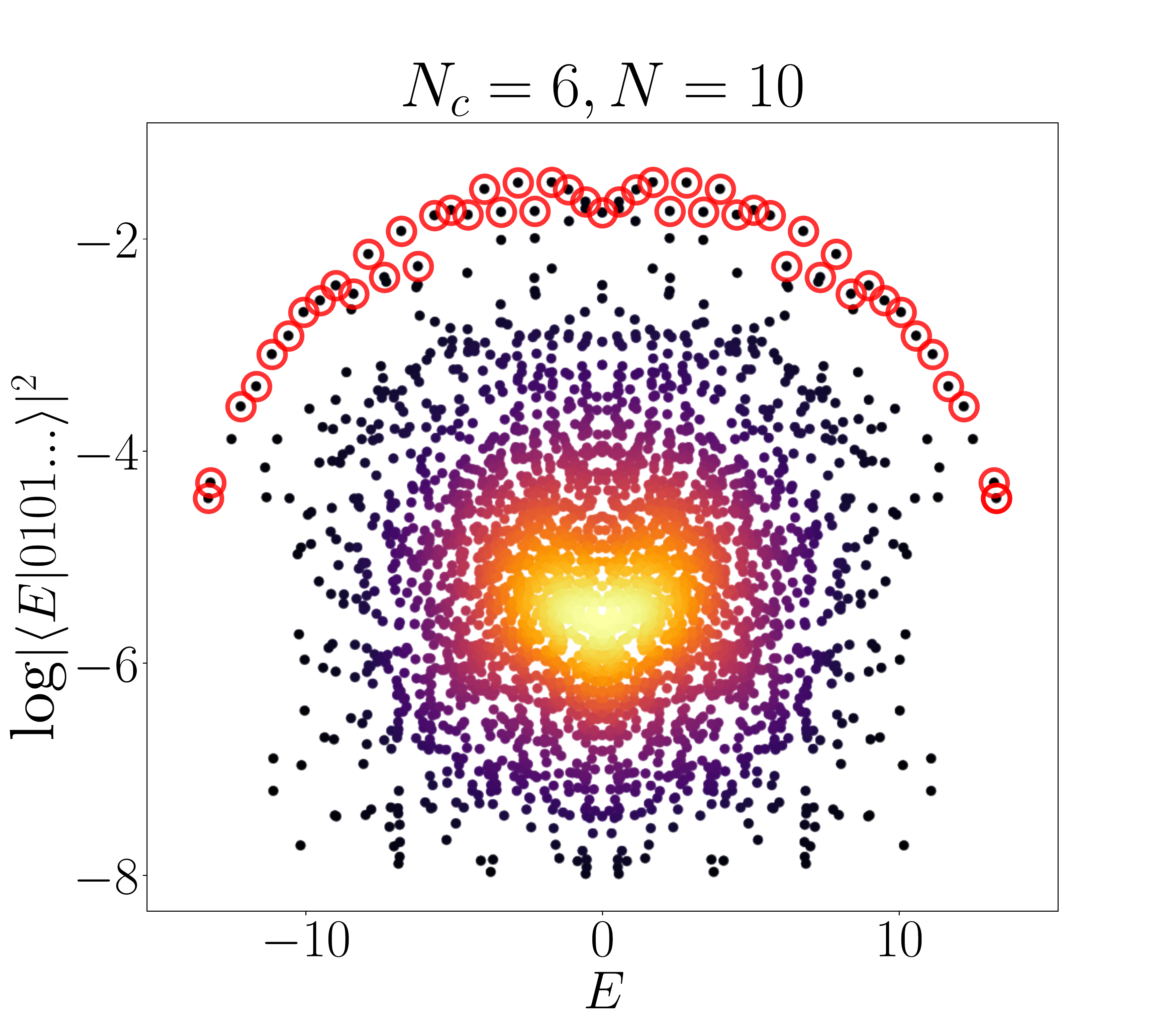}
\includegraphics[width=0.48\linewidth]{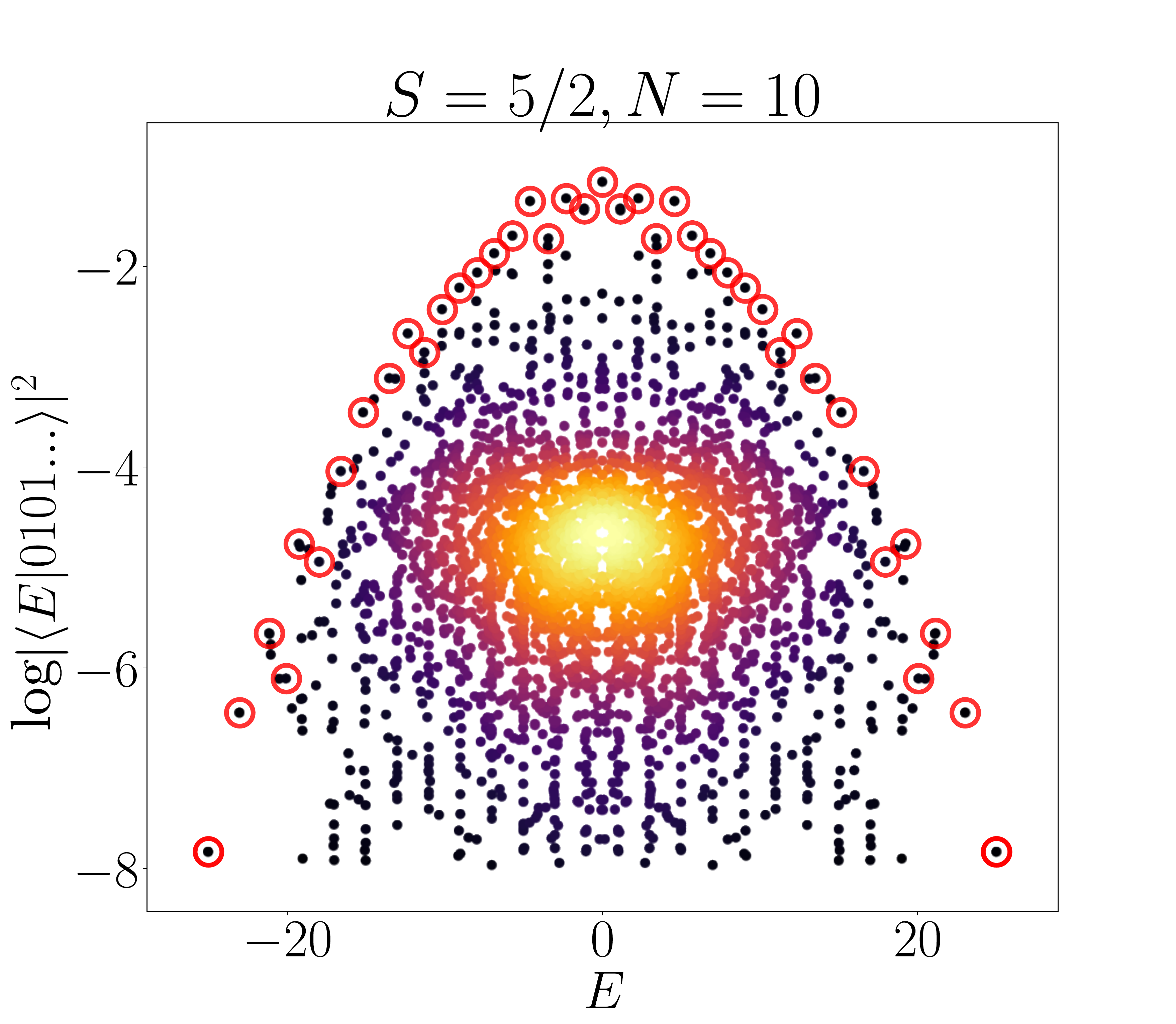}

\caption{Comparison of the scarred band of states for 5-color clock model and PXP spin-2 model. Each plot shows overlap of all eigenstates with the product state $|0404\ldots\rangle$. While the total number of scarred states is the same in the two models, the shape of the scarred band is quite different.}
\label{fig:5color}
\end{figure}

In Fig.~\ref{fig:5color} we compare the 5-color clock with the PXP spin-2 model and the 6-color clock with the PXP spin $5/2$ model. The different choice of the projector ($P^0$ or $P^{'}$) leads to a rather different structure of the scar states near the edge of the spectrum. While the corresponding clock and spin models both contain the same number of scar states, we see the band of scars remains better separated from the bulk near the edge of the spectrum in the clock models. Fidelity decays similarly in both models, but the spin model lacks the period of free precession observed in clock models in the main text, instead just bouncing back and forth between N\'eel and anti-N\'eel states, $\vert 0404... \rangle$ and $\vert 4040...\rangle$, as in spin-$1/2$ PXP. Clock models were found to ``shed'' less fidelity for odd $N_c$, where the difference in the first revival peak maximum of $\vert \langle \psi(t) \vert \psi(0) \rangle \vert^2$ was found to be 0.226 for $N_c=5$. For even $N_c$, however, the maximum fidelity is comparable between clock and spin models.

\section{Relation to chiral clock models}

By analogy with the quantum Ising model in a transverse field, Fendley has studied generalizations of models that involve clock operators~\cite{FendleyPara}, originally introduced by Baxter~\cite{Baxter1989}. Instead of focusing on critical properties of such models, we mainly focus on their fixed point,  where only the shift operator is present in the Hamiltonian. The first non-trivial case that is distinct from our clock operators occurs for $N_c=4$ Chiral Clock Model (CCM)~\cite{FendleyPara}, with the Hamiltonian defined as:
\begin{eqnarray}\label{eq:chiralclock}
    H = f \sum_{j=1}^N \tau_j^{\dagger} e^{-i \phi} + J \sum_{j=1}^{N-1} \sigma_j^{\dagger} \sigma_{j+1} e^{-i \theta} + h.c,
\end{eqnarray}
 where the operators $\sigma$ and $\tau$ generalize the spin-1/2 Pauli matrices, $\sigma_z$ and $\sigma_x$, respectively. They are given by
\begin{eqnarray}
    \sigma_j = \begin{pmatrix} 1 & 0 & 0 & 0 \\ 0 & \omega & 0 & 0 \\ 0 & 0 & \omega^2 & 0  \\ 0 & 0 & 0 & \omega^3 \end{pmatrix}, \;\;\;  \;\;\;
    \tau_j = \begin{pmatrix} 0 & 1 & 0 & 0 \\ 0 & 0 & 1 & 0 \\ 0 & 0 & 0 & 1 \\ 1 & 0 & 0 & 0 \end{pmatrix},
\end{eqnarray}
where (for $N_c=4$) we have $\omega = i$. Taking $J=0$, $\phi=-\pi / 2$, the Hamiltonian (\ref{eq:chiralclock}) becomes non-interacting, with the on-site Hamiltonian reducing to:
\begin{eqnarray}\label{eq:hsite}
    H^{site} = \begin{pmatrix}  0 & -i & 0 & i \\ i & 0 & -i & 0 \\ 0 & i & 0 & -i \\ -i & 0 & i & 0 \end{pmatrix}.
\end{eqnarray}
We construct an interacting many-clock model by introducing the same Rydberg-like constraint:
\begin{align}\label{eq:cc}
    H_{chiral} = \sum_{j} P_{j-1} H^{site}_j P_{j+1},
\end{align} 
with $P_j = |0_j\rangle\langle 0_j|$. This Hamiltonian can be obtained as the deformation of our clock model introduced in the main text.

\begin{figure}[htb]
\includegraphics[width=\linewidth]{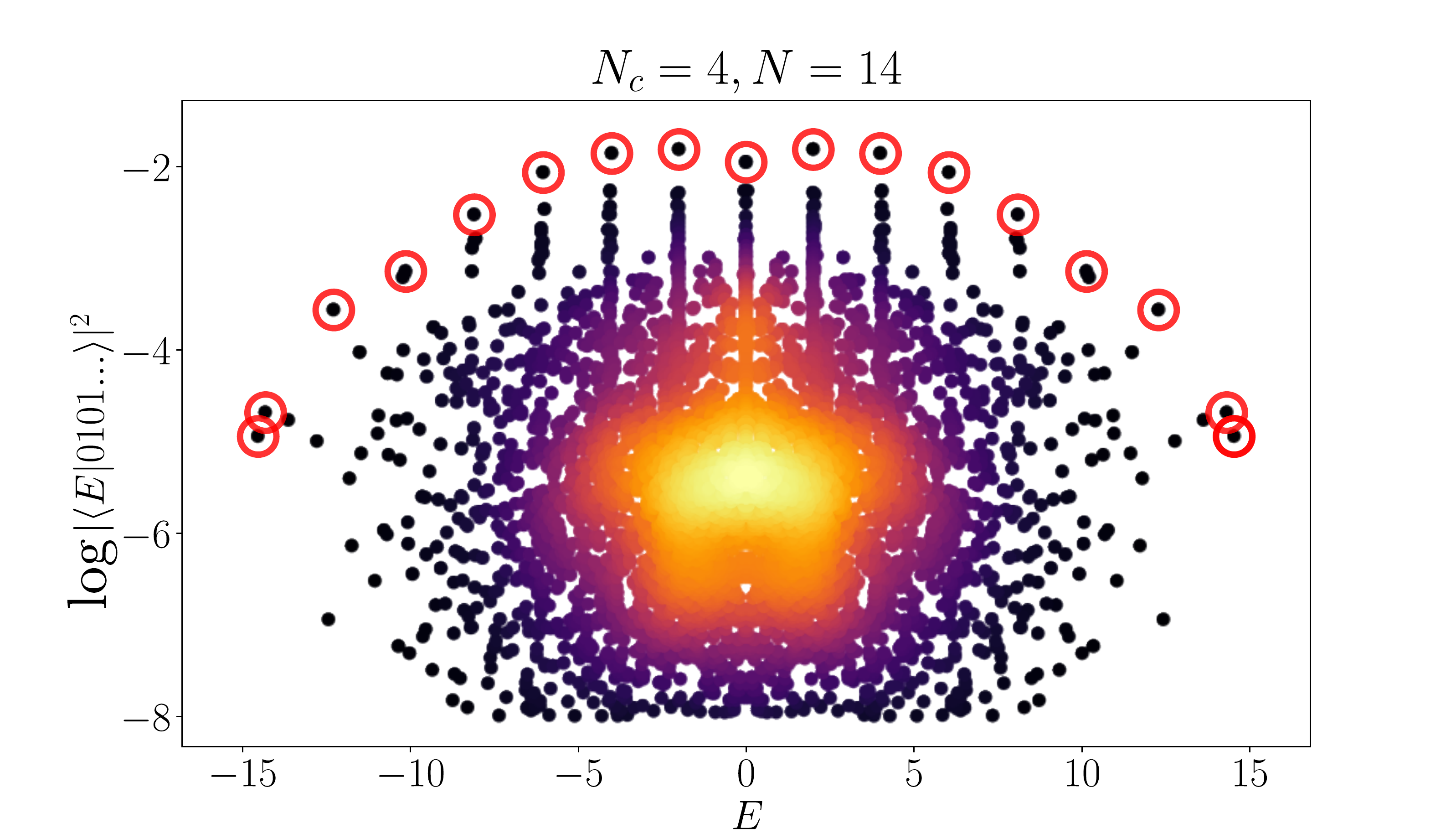}
\includegraphics[width=\linewidth]{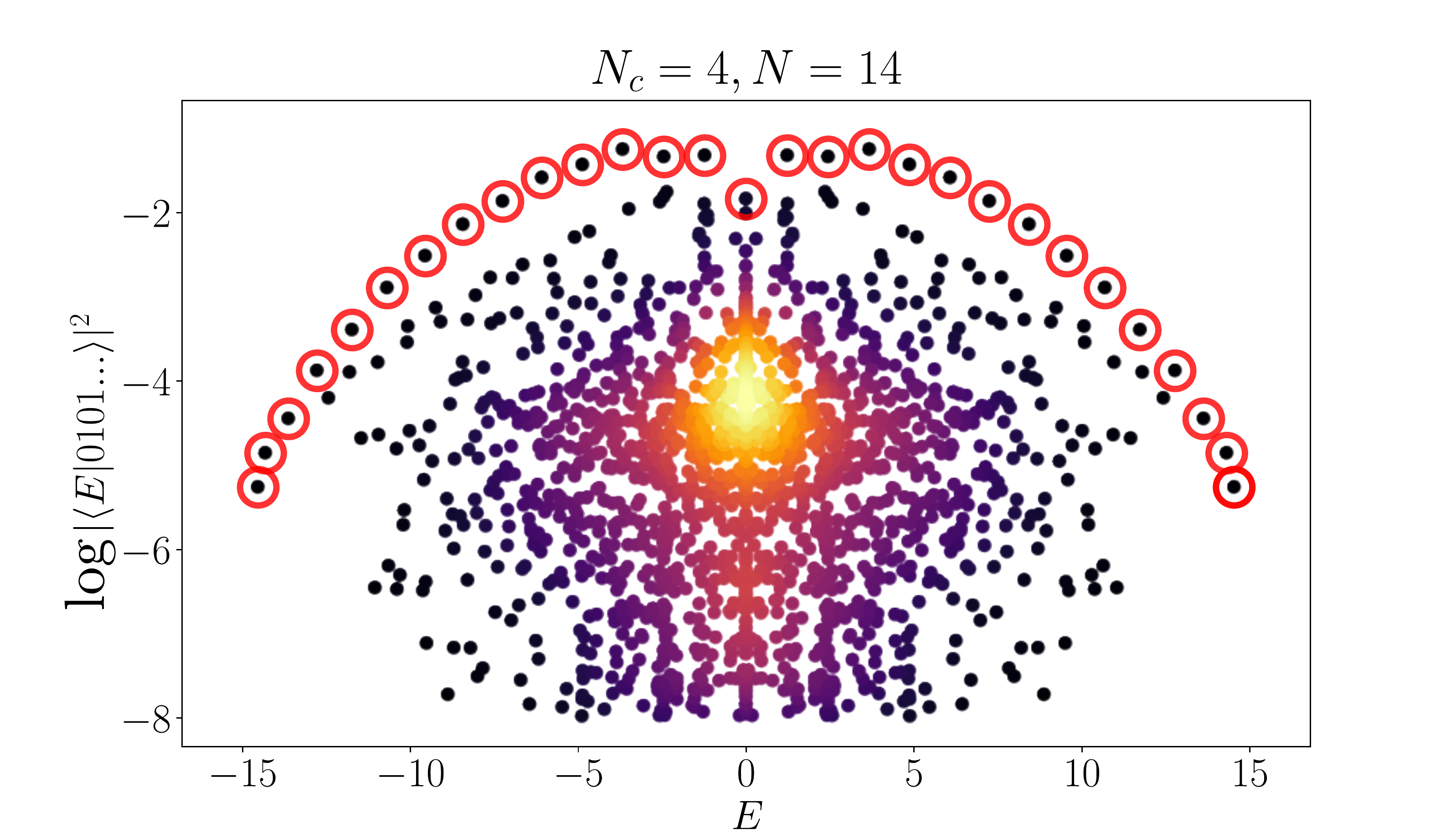}
\caption{Overlap of all eigenstates of CCM, Eq.~(\ref{eq:cc}), with product state $\vert0101\ldots\rangle$ (top) and $\vert0202\ldots\rangle$ (bottom). System size $N=14$. $\vert0101...\rangle$ has $N+1$ scars, while $\vert0202...\rangle$ has $N+2$ scars. }
\label{fig:chiralclock_overlap}
\end{figure}
In Fig.~\ref{fig:chiralclock_overlap} we plot the overlap of all eigenstates of the Hamiltonian in Eq.~(\ref{eq:cc}) with two product states states: $|1010\ldots\rangle$ (top) and $|2020\ldots\rangle$ (bottom). System size is $N=14$ with periodic boundary conditions.  These plots show that the model is not completely ergodic and supports a band of scarred states. However, the scarred bands look rather different in the two cases, with a stronger clustering of eigenstates into towers for the case of $|1010\ldots\rangle$ state. We note that the structure of the scarred band in the bottom panel of Fig.~\ref{fig:chiralclock_overlap} is quite reminiscent of the PXP spin-1/2 model~\cite{TurnerPRB}.  In fact, as we show next, the dynamics in this case reduces to the same type of oscillation that was found in the PXP spin-1/2 model.

\begin{figure}[h]
    \centering
    \includegraphics[width=0.5\textwidth]{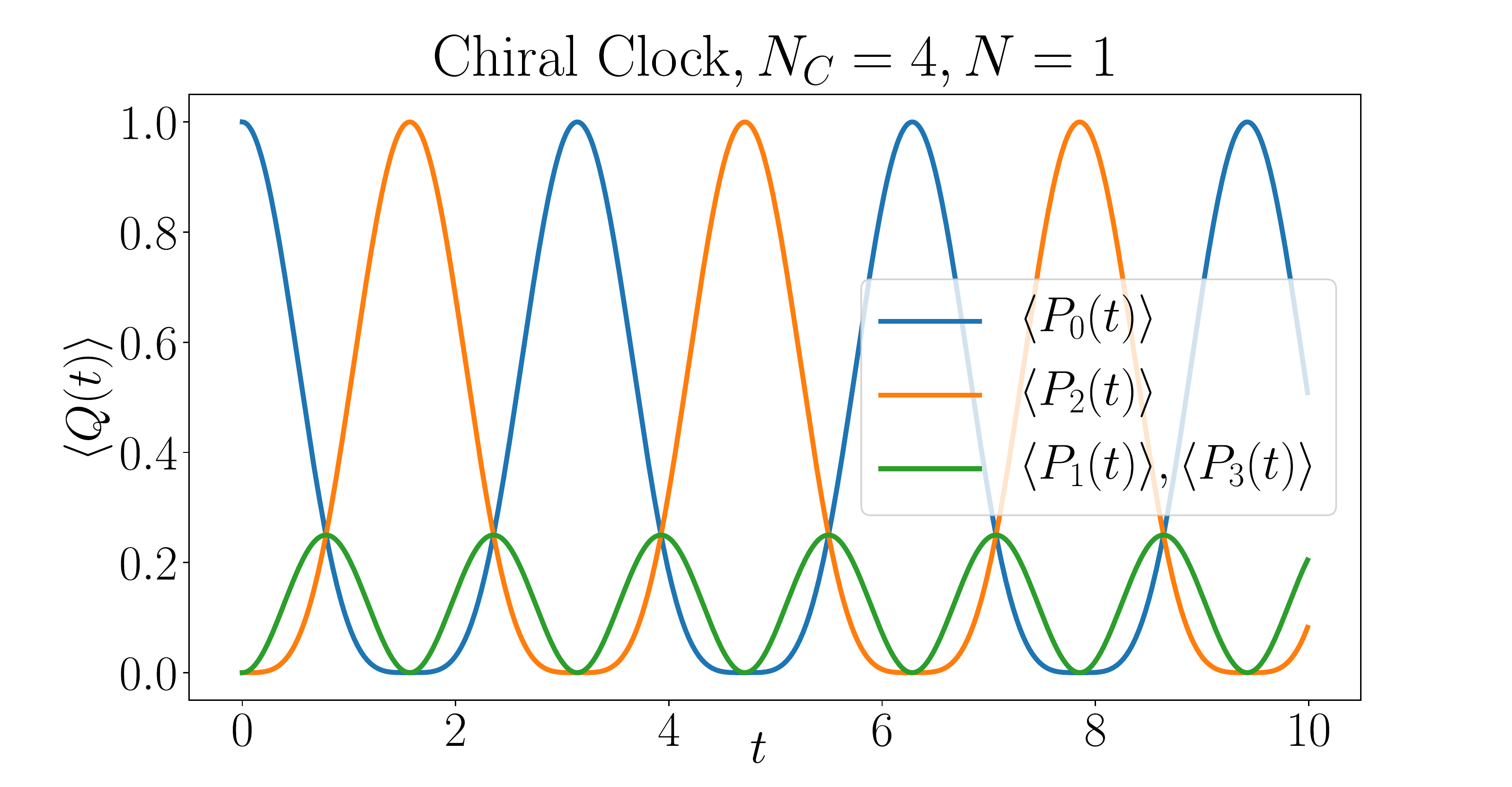}
    \includegraphics[width=0.5\textwidth]{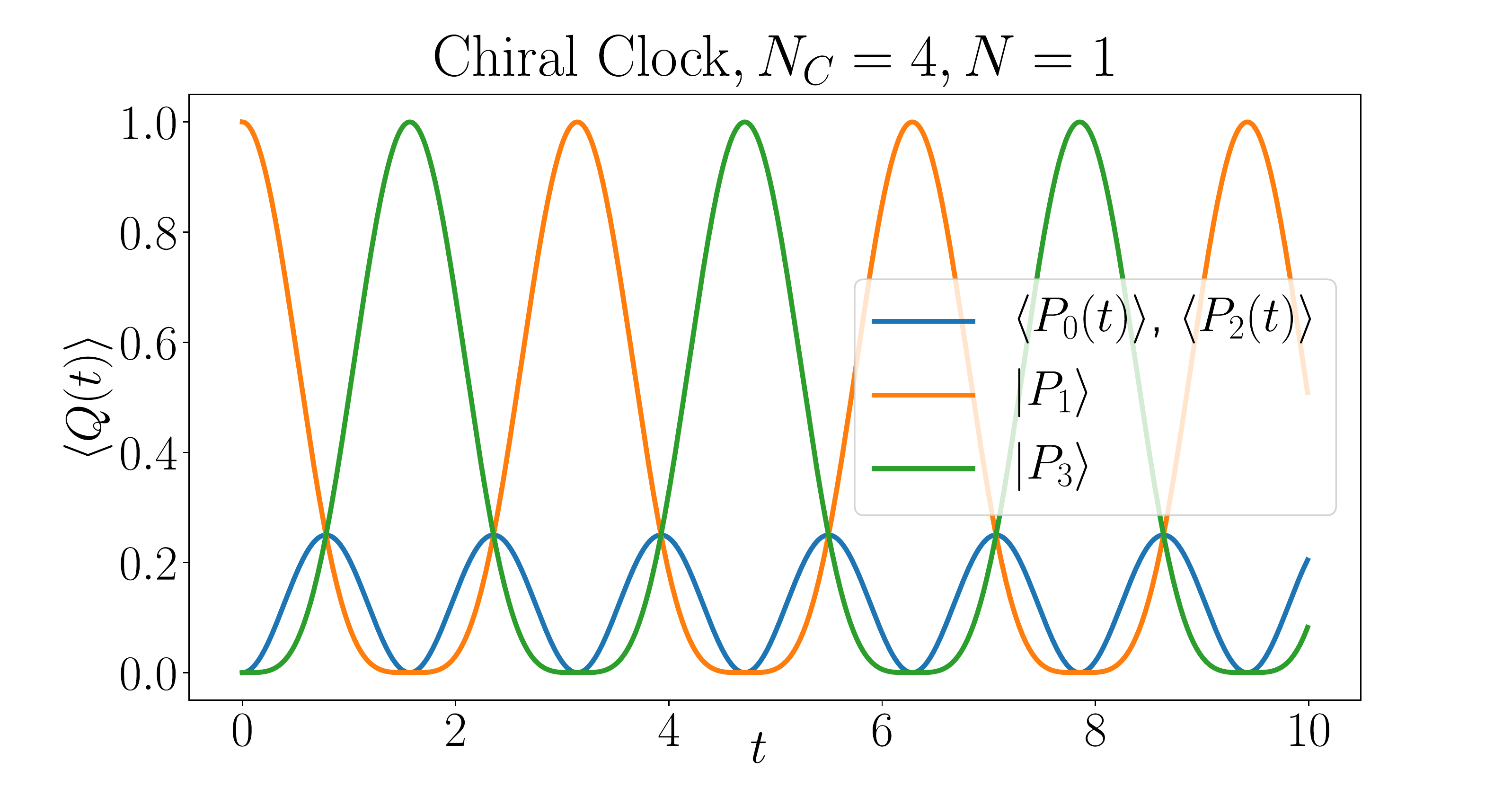}
    \caption{Single site precession of the unconstrained $H_{site}$ model in Eq.~(\ref{eq:hsite}). The dominant dynamics is $0\rightarrow2$ flips and $1\rightarrow3$ flips. } 
    \label{fig:chiral_free}
\end{figure}
Fig. \ref{fig:chiral_free} shows the single site precession of $H^{site}$ in the absence of a constraint.  We see that the dominant dynamics is $0\rightarrow2$ flips and $1\rightarrow3$ flips. Thus, when the constrained model, dressed with $P^0$, is quenched from $\vert 0101...\rangle$, the dominant dynamics is just to oscillate between $\vert 0101...\rangle \rightarrow \vert0303...\rangle$, as $0\rightarrow2$ is blocked due to the constraint. Neglecting leakage, this is just a decoupled free paramagnet precessing at every other site. However, when quenched from $\vert 0202...\rangle$, a transition through the polarized state $\vert 0000...\rangle$ is permitted and we see transitions like $\vert 0202 ... \rangle \rightarrow \vert 2020... \rangle$, reminiscent of spin-$1/2$ PXP.

\begin{figure}[htb]
\includegraphics[width=\linewidth]{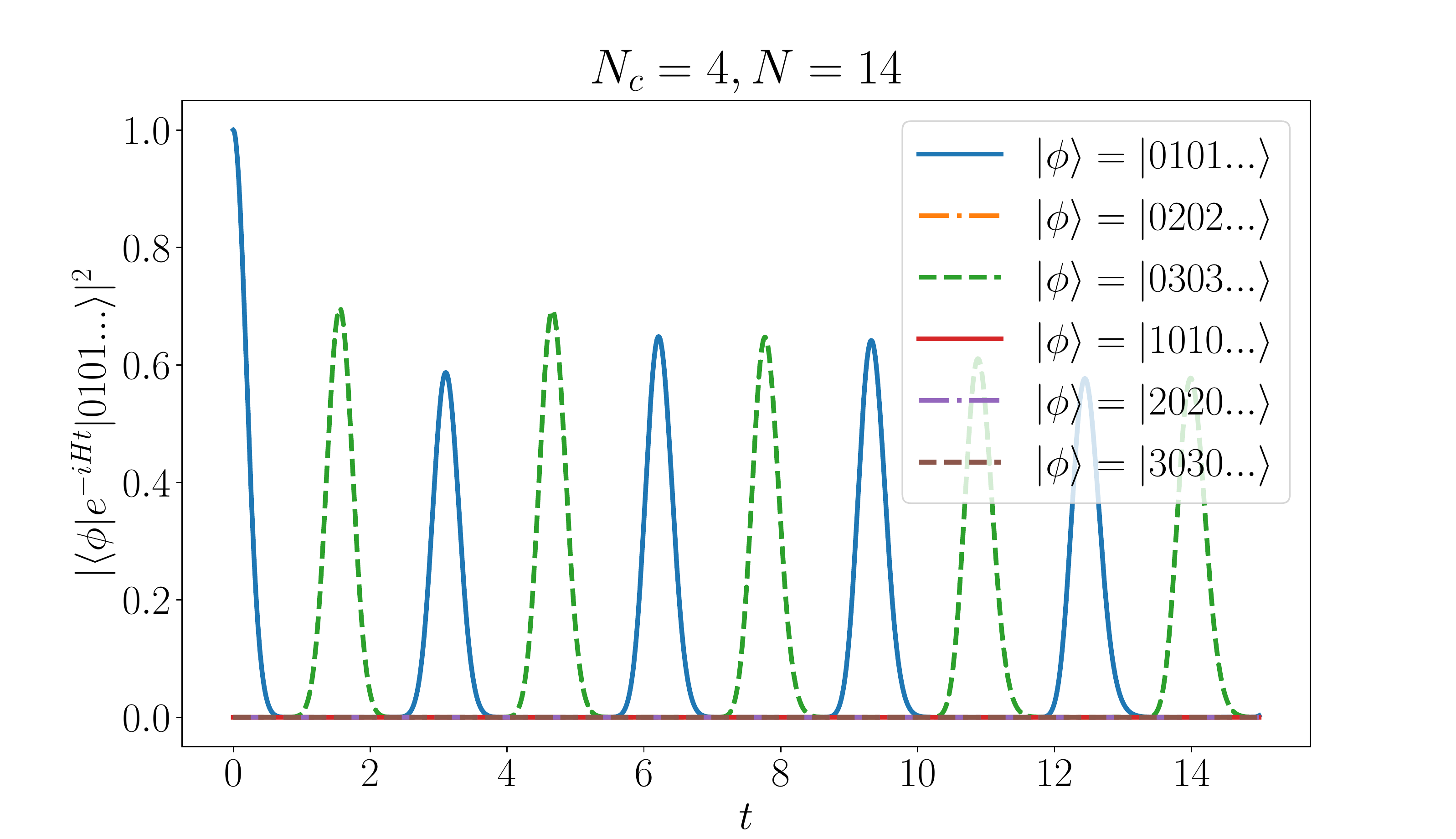}
\includegraphics[width=\linewidth]{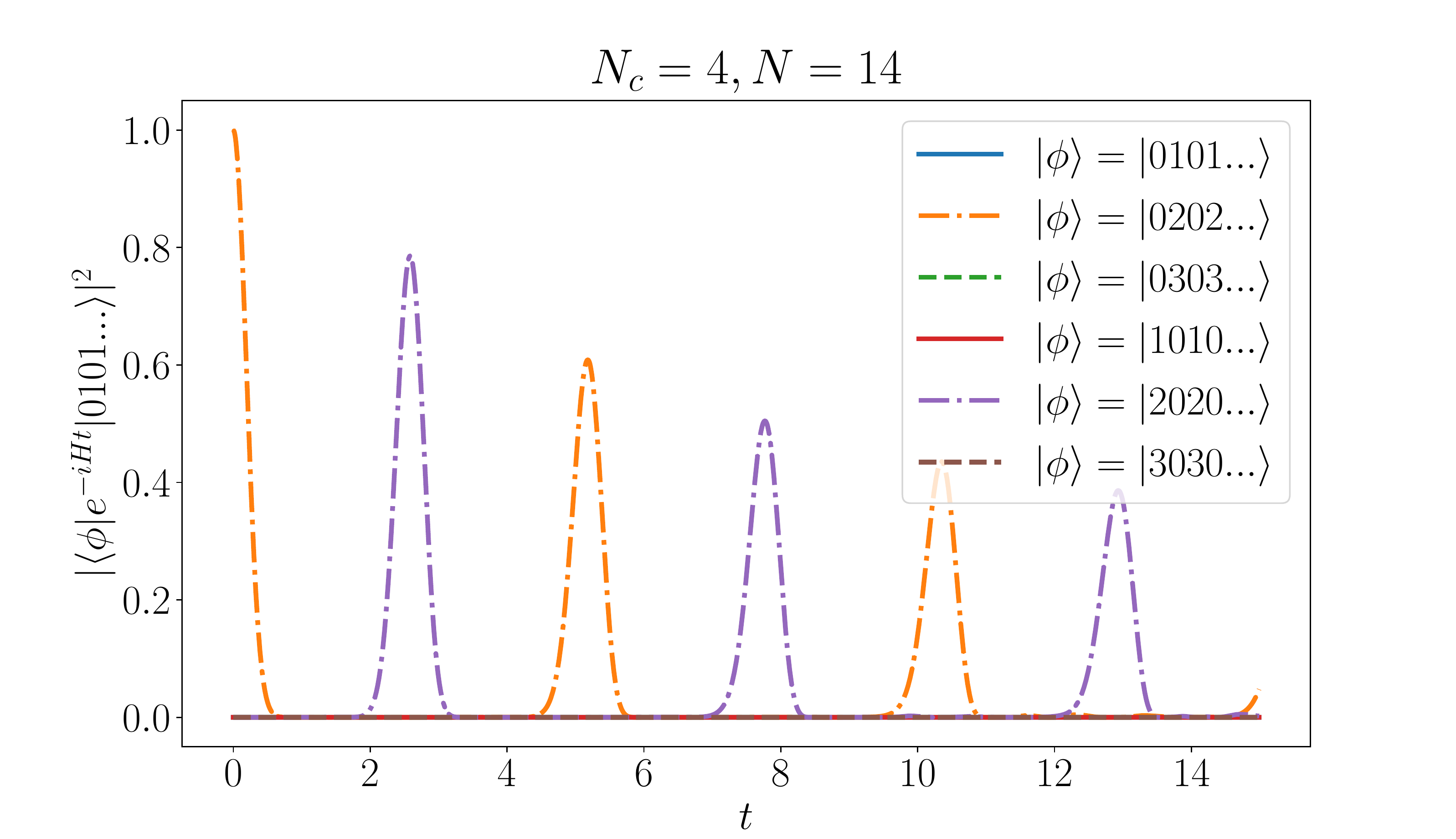}
\caption{Fidelity of revivals in CCM, Eq.~(\ref{eq:chiralclock}),  from the initial states $|1010\ldots\rangle$ (top) and $|2020\ldots\rangle$ (bottom). System size $N=14$. The plots show $|\langle \phi | \exp(-i t H) | \psi_0\rangle|^2$, where for $|\phi\rangle$ is any  of $|n0n0\ldots\rangle$, $n=1,2,3$, and their translated copies, $|0n0n\ldots\rangle$. 
}
\label{fig:chiralclock_fidelity}
\end{figure}
We confirm this by studying the fidelity time series, shown in Fig.~\ref{fig:chiralclock_fidelity}, for two initial states, $|\psi_0\rangle = |1010\ldots\rangle$ (top) and $|\psi_0\rangle =|2020\ldots\rangle$. As in the main text, to get a clearer insight into the dynamics, we plot the generalized fidelity, $|\langle \phi | \exp(-i t H) | \psi_0\rangle|^2$, where for $|\phi\rangle$ we do not necessarily pick the initial state, but any $|n0n0\ldots\rangle$, $n=1,2,3$, and their translated copies, $|0n0n\ldots\rangle$.  
We notice that the predominant  dynamics of this Hamiltonian is to drive oscillations between the $\vert 1 \rangle$, $\vert 3 \rangle$ states and the $\vert 0 \rangle$, $\vert 2 \rangle$ states, with a small leakage into the remaining states. When quenched from $\vert 0101...\rangle$, as $0\rightarrow2$ is blocked due to the Rydberg constraint, the  resulting dynamics is primarily just a free precession of the decoupled spin-$1/2$ chain, with a small leakage. This is observed in Fig. \ref{fig:chiralclock_fidelity}, where we see transitions between $\vert 0101...\rangle \rightarrow \vert 0303...\rangle$ without passing through $\vert0000....\rangle$. In contrast, $0\rightarrow2$ flips are permitted when quenching from the state $|2020\ldots \rangle$. This allows the state to evolve through the polarized state $\vert 0000...\rangle$ and exchange sublattices, undergoing oscillations between $\vert 0202...\rangle \rightarrow \vert 2020...\rangle$. This is precisely the same dynamics of spin-$1/2$ PXP, neglecting the leakage. As we see in Fig.~\ref{fig:chiralclock_fidelity} (bottom), we obtain non-zero fidelity only with the states $|2020\ldots\rangle$ and $|0202\ldots\rangle$, suggesting that the 4-color chiral clock model contains an embedded PXP spin-1/2. 

Finally, we note that CCMs, as defined in Ref.~\cite{FendleyPara}, appear to have scars only in $N_c\leq 4$ cases. For odd  $N_c \geq 5$, we did not find revivals from N\'eel-type initial states. However, we note that for $N_c$-even it is possible to construct a special case of our clock models by choosing $k_n \in \{-N_c/2,...,-1,1,...N_c/2\}$. These models \emph{do} appear to support scars, and for $N_c=4$  this choice of $k$'s reduces to CCM in Eq.~(\ref{eq:chiralclock}). Nevertheless, for  $N_c>4$ longer range couplings are introduced and the models are no longer equivalent to Ref.~\cite{FendleyPara}. The scarring in these models can similarly be explained by studying the single-site free precession, where dominant cyclic transitions between basis states emerge, with cycles shorter than $N_c$. Ignoring leakage away from these paths, they are effectively the same as the corresponding clock model with a matching cycle length. 

\section{Analytic prediction of parameters of 4-color scarred models}

In the main text we presented a phase diagram of $N_c=4$ models of the form $H=\sum_n P^{0}_{n-1} C_n P^{0}_{n+1}$, which were shown to contain strong revivals. We noted that optimal models in these diagrams could be roughly predicted by determining the values of parameters that yield a single-site Hamiltonian $C$ with evenly spaced eigenvalues, such that the single-site dynamics is periodic. Further models with weaker revivals are obtained for  $C$ whose eigenvalue spacings differed by a ratio of 2. In the following we derive analytical expressions for the lines in Fig.~2 in the main text.

Varying the matrix elements of $C$, we consider all Hermitian, purely imaginary, non-diagonal $C$, in analogy with the clock and spin Hamiltonians discussed in the main text. We are free to rescale $C$ such that one coupling has magnitude $1$, so for $N_c=4$, $5$ couplings can be varied. We consider two slices of the phase diagram: (A) vary the next-nearest neighbor hoppings, $C_{02} = C_{13} = \alpha i$, while also varying $C_{03} = -\beta i$; (B) switch off next-nearest-neighbor hoppings while varying $C_{12} = -\alpha i$ and $C_{03} = -\beta i$. Explicitly:
\begin{align}
    C_n^{(A)} &= 
    \begin{pmatrix}
        0 & -i & \alpha i & -\beta i \\
        i & 0 & -i & \alpha i \\
        -\alpha i & i & 0 & -i \\
        \beta i & -\alpha i & i & 0\\
    \end{pmatrix}, \label{eq:c_typeA}\\
    C_n^{(B)} &= 
    \begin{pmatrix}
        0 & -i & 0 & -\beta i \\
        i & 0 & -\alpha i & 0 \\
        0 & \alpha i & 0 & -i \\
        \beta i & 0 & i & 0 \\
    \end{pmatrix}. \label{eq:c_typeB}
\end{align}

\begin{figure}[htb]
    \centering
    \includegraphics[width=0.5\textwidth]{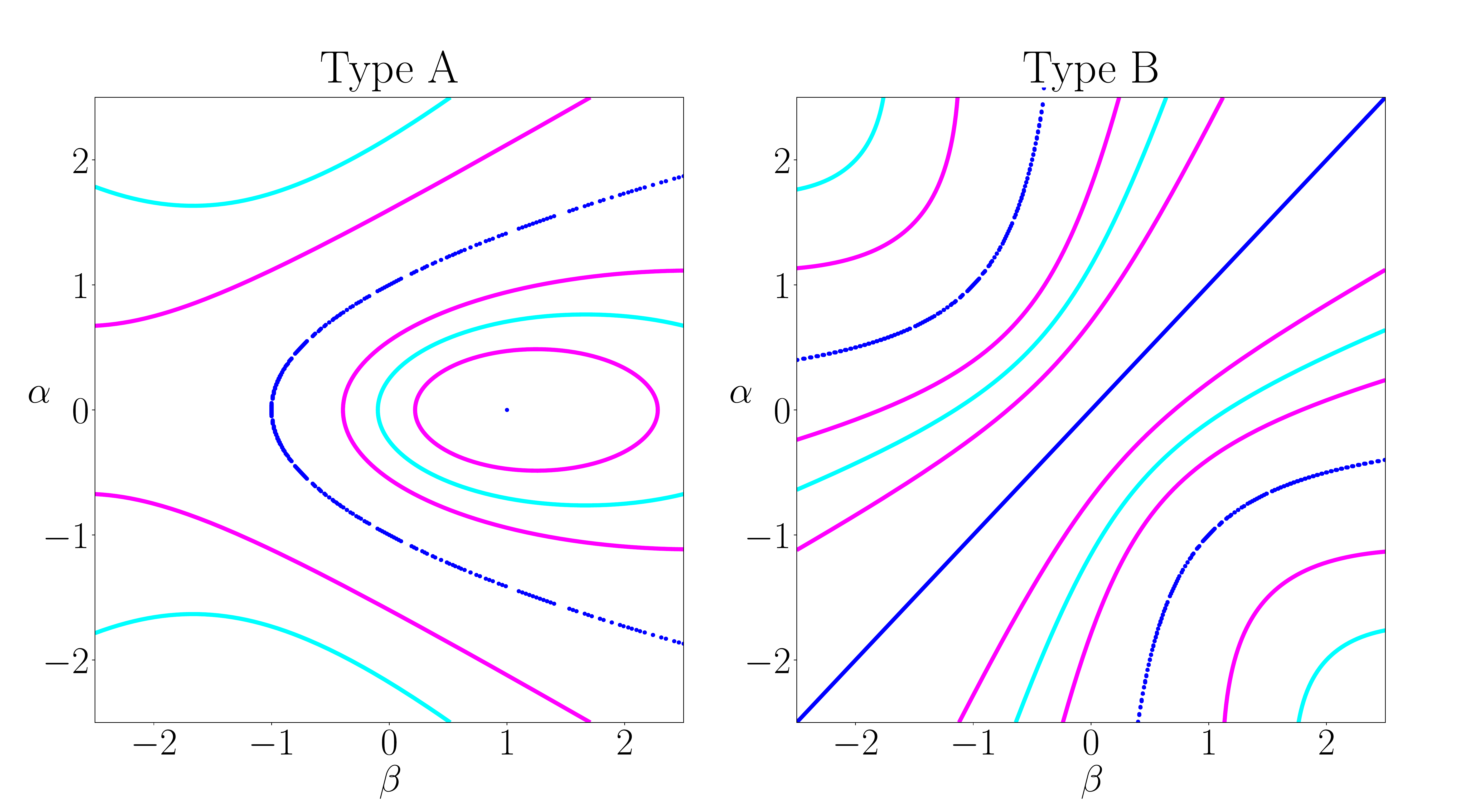}
    \caption{Prediction of scarred models with strongest revivals based on the eigenvalue spacings of $C$ for $N_c=4$. Left and right panels refer to two deformations of  $C$ in Eqs.~\ref{eq:c_typeA}, \ref{eq:c_typeB}. Plotted here are: (a) equidistant eigenvalues, Eq.~(\ref{eq:even_energies}) (cyan), (b) eigenvalue spacings commensurate with ratio 2 (pink), Eq.~(\ref{eq:comm_two_a}), (\ref{eq:comm_two_b}), (c) degenerate eigenvalues, Eq.~(\ref{eq:degen_a}), (\ref{eq:degen_b}) (blue). These lines coincide with the regions of strong revivals in Fig.~2 in the main text.} 
    \label{fig:analytic_phase}
\end{figure}
The characteristic eigenvalue equation for both these matrices reduces to the form
\begin{align}
    E^4-a E^2 + b = 0,
\end{align}
with $a$ and $b$ being:
\begin{align}
    \textrm{(A)} \quad a &= 3+2\alpha^2+\beta, \\
    b  &= 2 \beta - 2 \alpha^2 - 2 \alpha^2 \beta + \alpha^4 + \beta^2+1, \\
    \textrm{(B)} \quad a &= \alpha^2 + \beta^2 + 2, \\
    b  &= 2 \alpha \beta + \alpha^2 \beta^2 + 1. 
\end{align}
It follows that the eigenvalues $E$ take the form
\begin{align*}
    E_{1,\ldots,4} = \mp \frac{1}{\sqrt{2}} \sqrt{a \pm \sqrt{a^2-4b}}.
\end{align*}
These eigenvalues are symmetric about zero, such that the spacings $\Delta E_1 = \Delta E_3$, with $\Delta E_n = E_{n} - E_{n-1}$. The analytic prediction for robust scarred models, based on the single-site analysis, then reads:
\begin{itemize}
   \item  Equidistant eigenvalues:
\begin{eqnarray}
9a^2 = 100b, \quad \quad \Delta E_2 = \Delta E_1, \label{eq:even_energies} 
\end{eqnarray}   

\item Commensurate eigenvalues with ratio 2:
\begin{eqnarray}
&& 4a^2 = 25b, \quad \quad \Delta E_2 = 2 \Delta E_1, \label{eq:comm_two_a} \\
&& 25^2 = 676b, \quad \quad \Delta E_2 = \frac{1}{2} \Delta E_1,  \label{eq:comm_two_b} 
\end{eqnarray}

\item Degenerate eigenvalues:
\begin{eqnarray}
&& a^2 = 4b, \quad \quad \Delta E_1 = \Delta E_3 = 0, \label{eq:degen_a} \\
&& b=0, \quad \quad  \Delta E_2=0  \label{eq:degen_b}
\end{eqnarray}
   \end{itemize}
These equations are plotted in terms of parameters $\alpha, \beta$ in Fig~\ref{fig:analytic_phase}. These lines give the band of scarred models in our $N_c=4$ phase diagram in Fig.~2 in the main text.

\section{Open Boundary Conditions}

In the PXP spin-1/2 model~\cite{Turner2017,TurnerPRB}, open boundary conditions have little effect on the scarring physics. Special scarred eigenstates still exist in the open PXP chain, and oscillatory quench dynamics still follow from certain initial states (N\'eel), while rapid thermalization is observed from generic quenches. Fig~\ref{fig:obc} summarizes the dynamics of the $N_c=3$ colored clock, for the largest accesible system size with open boundary conditions ($N=14, dim(\mathcal{H})=21845$). Numerically, we observe the same scarring phenomenology as with periodic boundary conditions, with the same pattern of transitions $\vert 0101...\rangle \rightarrow \vert 0202...\rangle \rightarrow \vert1010...\rangle$. The only difference, as in PXP spin-1/2, is that revivals decay slightly faster in an open chain. 
\begin{figure}[h]
    \centering
    \includegraphics[width=0.5\textwidth]{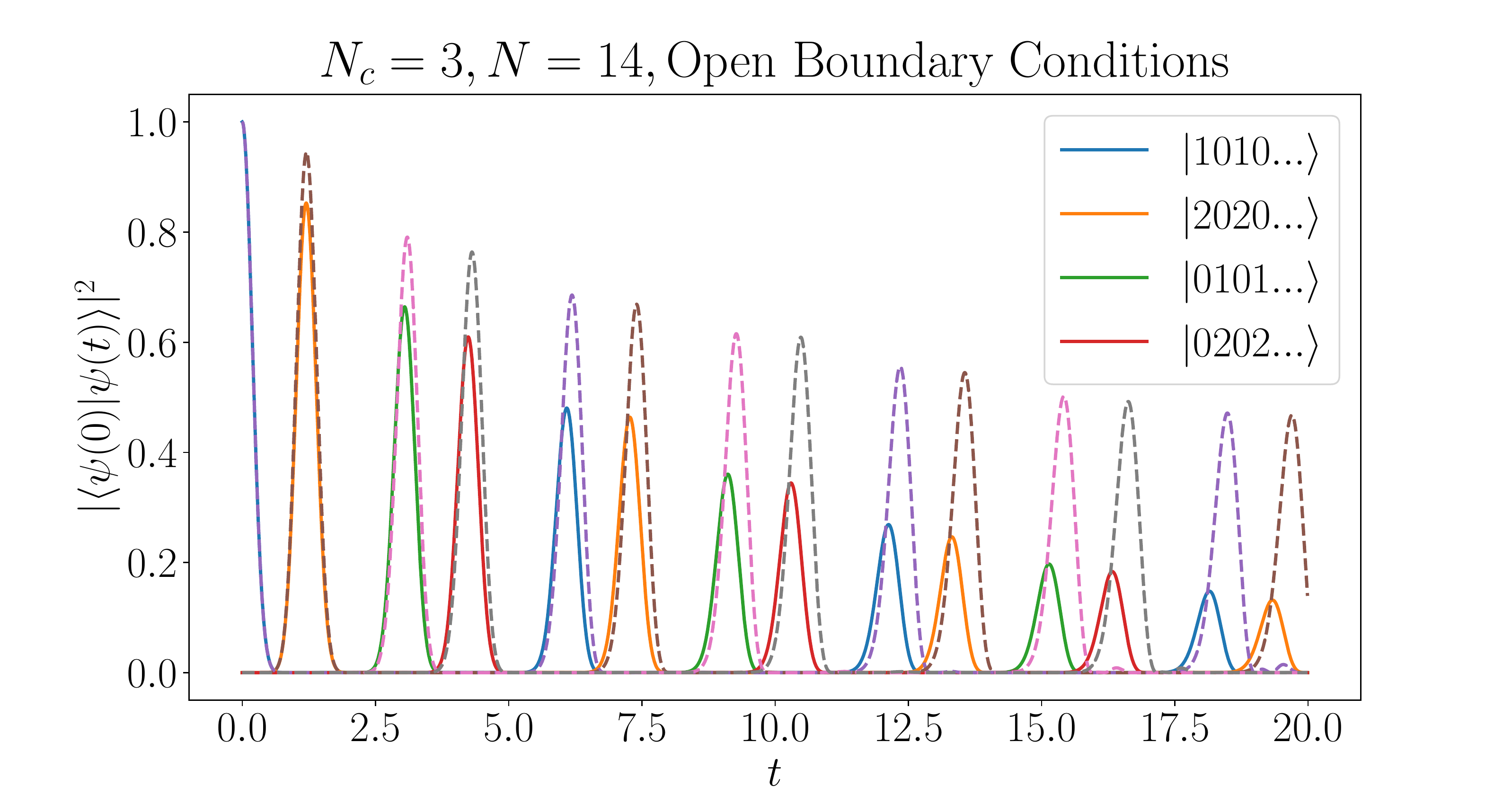}
    \includegraphics[width=0.5\textwidth]{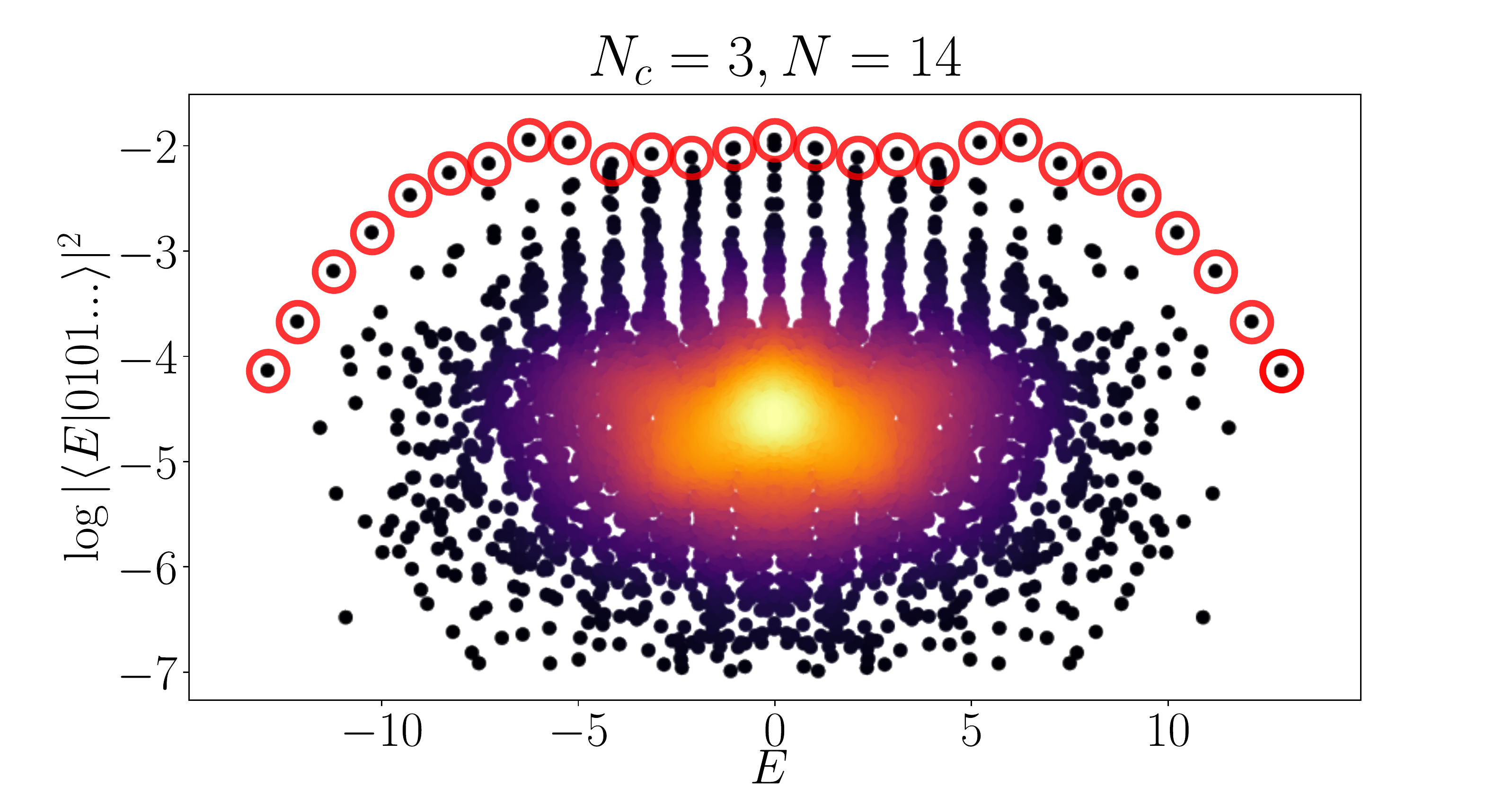}
    \caption{Dynamics of $N_c=3$ colored clock model with open boundary conditions. The same generic features exist as in the periodic case, including a period of nearly free precession followed by an interacting segment. }
    \label{fig:obc}
\end{figure}

\section{Finite Size Scaling of Numerical Results}

The system sizes accesible to us are quite limited, due to relying on  exact diagonalization techniques to obtain all eigenstates of systems with fairly large local Hilbert space dimensions. While our models could be studied using MPS/DMRG techniques~\cite{White,Vidal07}, we note such techniques can only be used to simulate dynamics at short times. By contrast, important information about the atypical eigenstates of the system are currently only obtainable via exact diagonalization methods.

Fig~\ref{fig:scaling} presents a finite size scaling analysis of our clock models, illustrating that our main results are well-converged in the accessible system sizes.  Bottom panel of Fig~\ref{fig:scaling} shows $t_0(N)$, the time at which the first maximum of the fidelity $F(t) = \vert \langle 1010... \vert e^{-iHt} \vert 1010...\rangle\vert^2$ occurs, while $f_0$ is the amplitude of this revival, $f_0=F(t_0)$. More precisely, top panel of Fig~\ref{fig:scaling} shows the fidelity density,  $\mathcal{F}(N) = \ln(f_0(N))/N$, which has a well-defined thermodynamic limit. The largest system sizes accesible to us were $N=20$ for $N_c=3$ and $N=16$ for $N_c=4$. Resolving both translational and parity symmetry, the Hilbert space dimension of the relevant sectors are of the order of $2\times 10^4$ for both $N_c=3$ and $N_c=4$. 
\begin{figure}[h]
    \centering
    \includegraphics[width=0.5\textwidth]{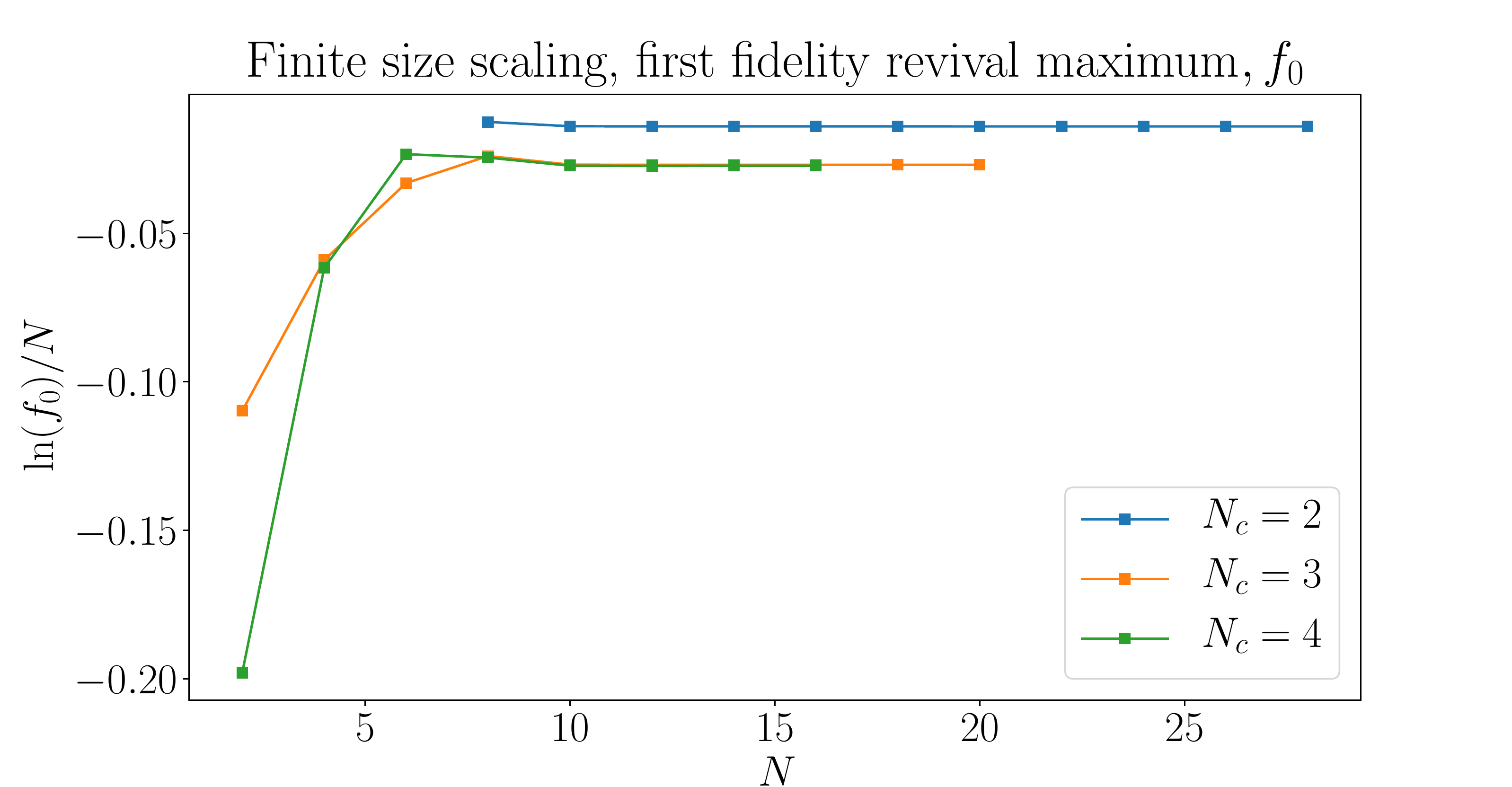}
    \includegraphics[width=0.5\textwidth]{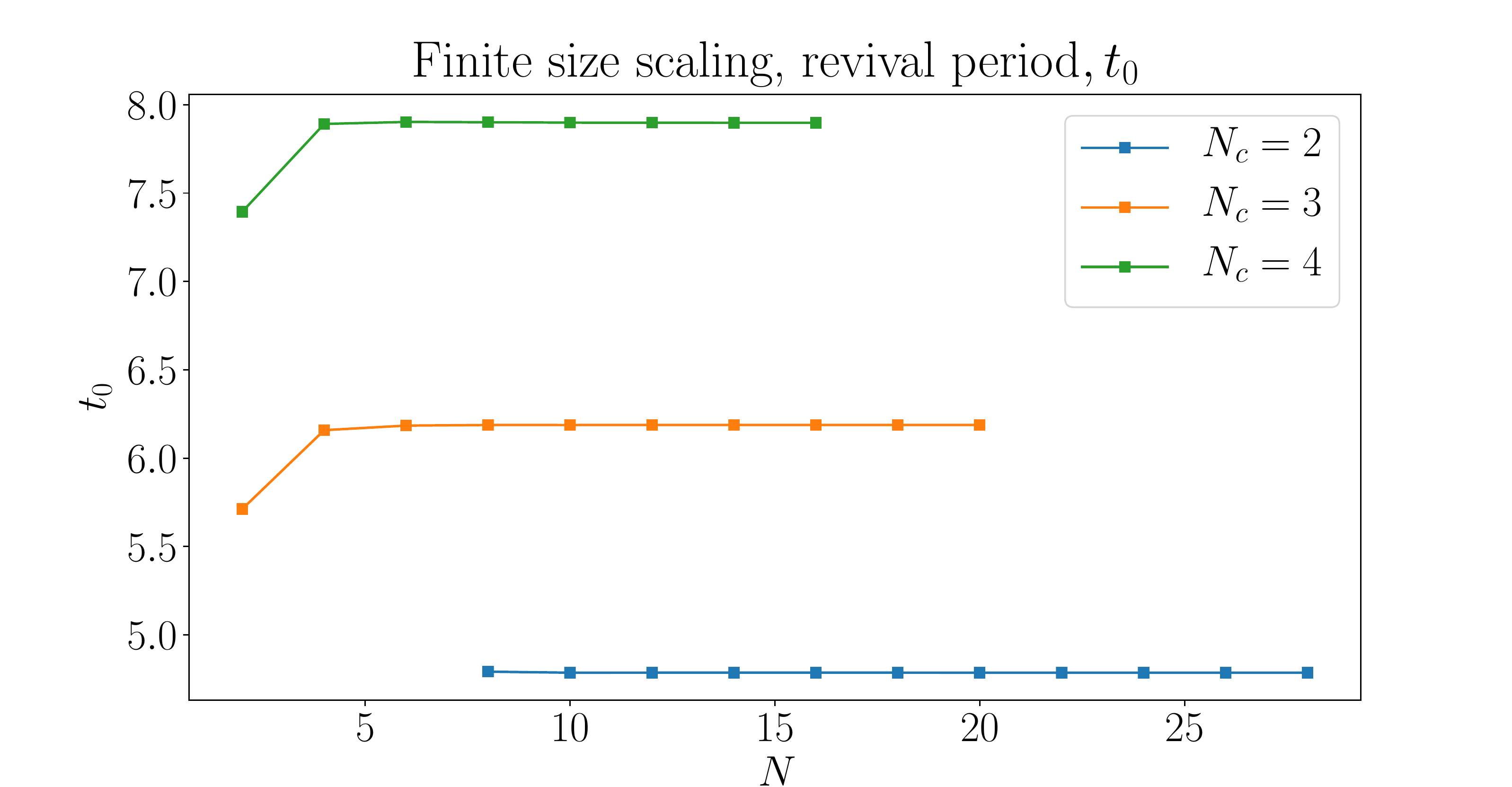}
    \caption{Finite size scaling analysis of revivals seen in clock models. The top plot considers $f_0(N)$, the first revival maximum of the fidelity $F=\vert \langle 1010... \vert e^{-iHt} \vert 1010...\rangle \vert^2$, while the bottom plot considers $t_0$, the time at which this first revival occurs.}
    \label{fig:scaling}
\end{figure}

\section{Approximating scarred eigenstates}

In the PXP spin-1/2 model~\cite{Turner2017,TurnerPRB} an approximation scheme called ``the forward scattering approximation" (FSA) was introduced in order to approximate  the $N+1$ scarred eigenstates in a chain of size $N$. This approximation  was shown to accurately capture the eigenstates that have highest overlaps with the N\'eel state and whose energies are equally spaced. The FSA is formally based on the splitting of  the Hamiltonian into two parts, $H=H^++H^-$, where $H^+$ increases the Hamming distance from the N\'eel state by one (and $H^-$, conversely, decreases this distance by one). The FSA is defined by projecting the PXP Hamiltonian into the $(N+1)$-dimensional basis spanned by vectors $(H^+)^n|1010\ldots\rangle$, $n=0,1,\ldots,N$. By construction, these vectors are orthogonal to each other. Diagonalizing the PXP Hamiltonian in this basis produces approximate eigenenergies and eigenvectors, which were found to match the exact ones with high accuracy. Moreover, the projection of the Hamiltonian into the FSA basis can be performed analytically, allowing one to reach much larger sizes than in exact diagonalization. Finally, the FSA scheme also provided an intuitive
picture for the quench dynamics in terms of the wave function spreading over the Hilbert space graph~\cite{Turner2017}. 

\begin{figure}[htb]
\includegraphics[width=\linewidth]{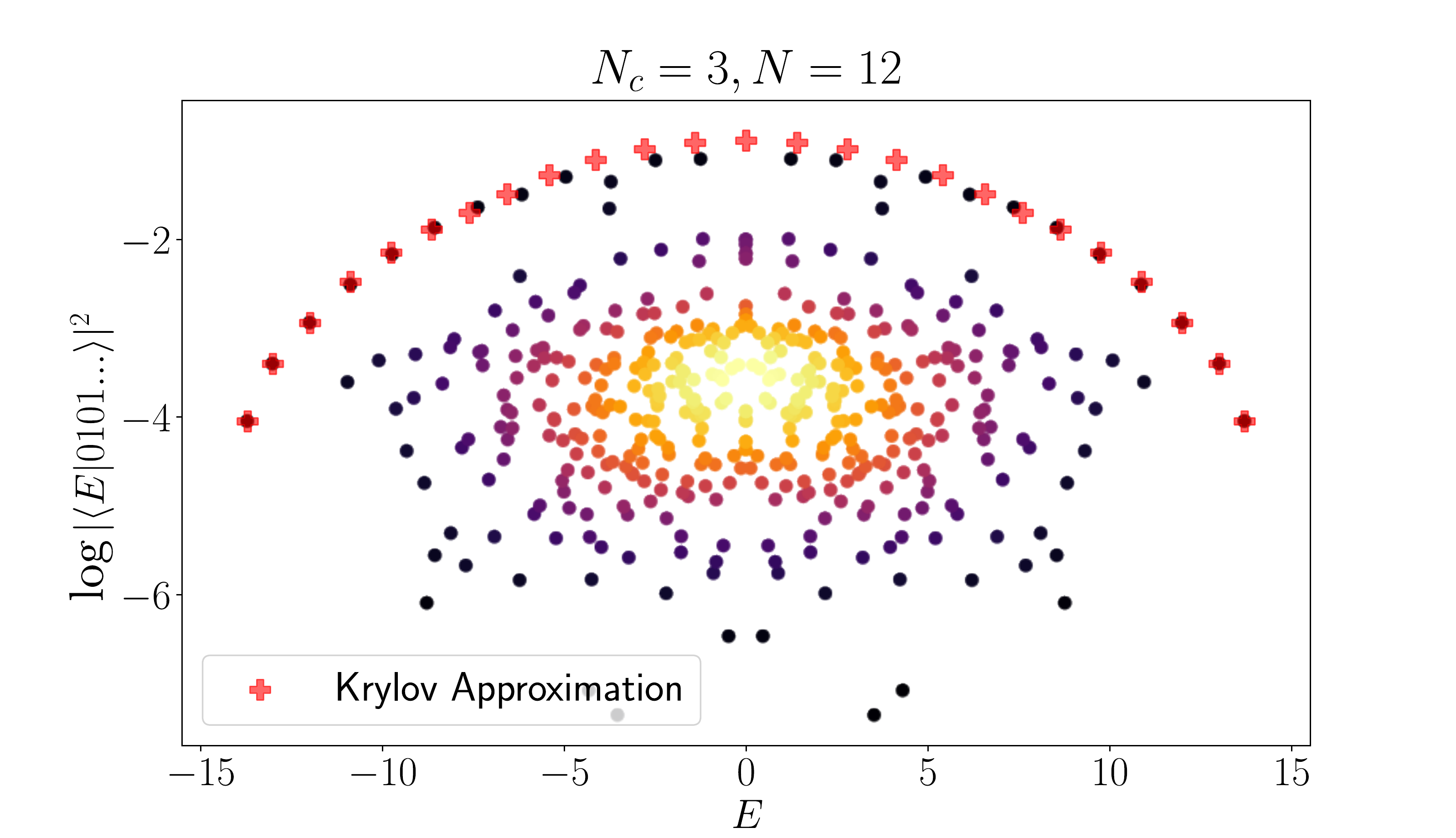}
\includegraphics[width=\linewidth]{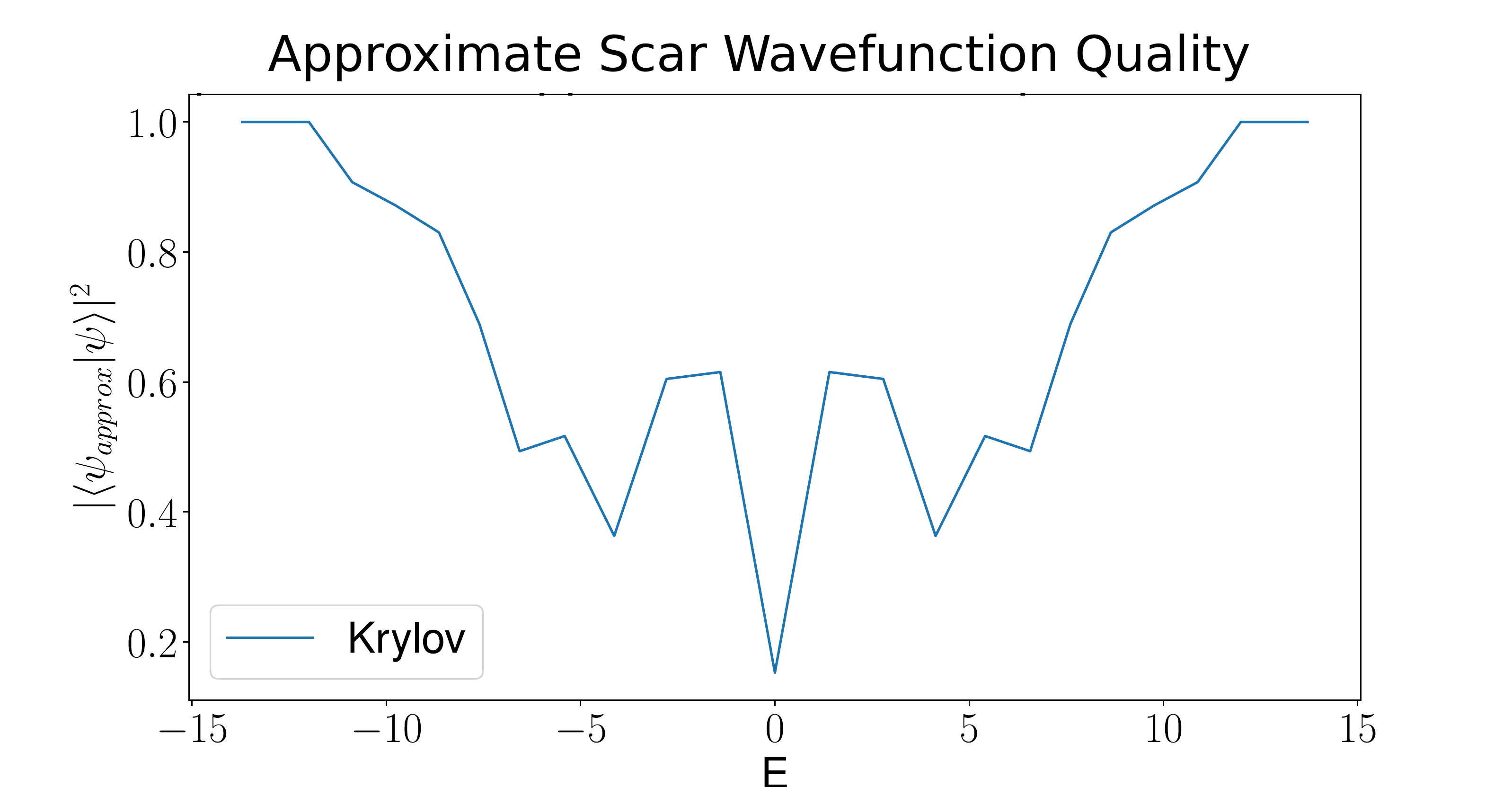}
\caption{Quality of scar approximations obtained by projecting $H$ into the $2N+1$ Krylov subspace, using $\vert 0101...$ as the initial vector.}
\label{fig:krylov}
\end{figure}
Unfortunately, direct generalization of the FSA to $N_c>2$ clock models is not trivial.
In retrospective, in the PXP case, the splitting of the Hamiltonian into $H^+$ and $H^-$ is straightforward because we can rely on the conventional definition of the Hamming distance. In the higher color models, this is no longer sufficient; moreover,  it may be necessary to split the Hamiltonian into more than two parts. Nevertheless, in Fig.~\ref{fig:krylov} we demonstrate that a more general approximation scheme yields accurate results for the $N_c=3$ clock model. Instead of splitting $H$, we follow the standard Lanczos procedure and build the Krylov subspace using full $H$: $\mathcal{K}=\{ H^n |1010\ldots\rangle,  n=0,1,2,\ldots \}$. Compared to the FSA, the Krylov subspace may contain an arbitrary number of vectors and we need to explicitly impose a cutoff. Furthermore, the Krylov vectors are no longer necessarily orthogonal, and Gram-Schmidt orthogonalization needs to be explicitly performed. We note that this approximation scheme is identical to the standard Lanczos algorithm for approximating the \emph{ground state} of a system.  In that case, however, the size of the Krylov subspace can be as large as as required to reach a target precision, while in our case we stop the algorithm after a given number of steps. Moreover, we obtain approximations to \emph{all} scarred states in a single run of the algorithm. 

Fig.~\ref{fig:krylov}(top) shows the result of the Krylov approximation for the 3-color clock model. We see that the approximation accurately predicts the energies of scarred eigenstates, as well as the eigenstate overlap with  $|1010\ldots\rangle$ product state. Fig.~\ref{fig:krylov}(bottom) also shows the overlap of the predicted scarred wave functions  with those obtained by exact diagonalization. The approximation is very accurate for the ground state, which is expected given that we are effectively using the Lanczos algorithm. Neglecting the zero energy state (which is not uniquely defined due to large degeneracy at $E=0$), the average overlap with the exact states for $N_c=3, N=12$ is $0.7410$. This is considerably smaller compared to PXP $1/2$, where the same procedure results in an average accuracy of $0.9173$. Moreover, the explicit orthogonalization that needs to be performed numerically restricts the Krylov approximation scheme  to much smaller system size compared to the FSA.

\section{Improving Revivals}

We noted in the main text that revivals are slightly more robust in PCP models over PXP models for odd $N_c$. Revivals can be substantially improved further, for any $N_c$, by the addition of certain perturbations, which we discuss here.

For PXP spin-1/2, it has previously been shown~\cite{Khemani2018,Choi2018} that adding the family of perturbations
\begin{eqnarray}\label{eq:pert1}
 H_d^{PPXP} = \sum_n P_{n-1} X_n P_{n+1}(P_{n-d}+P_{n+d})
\end{eqnarray}
 can substantially improve revivals. Following the above discussion on the FSA approximation, the Hamming operators $H^+,H^-$ approximately obey an su(2) algebra, in the sense that $[H^+,H^-]=2i H^Z + Q$, where $Q$ is the deviation from a perfect su(2)  algebra. One can show the perturbations $H_d^{PPXP}$ can reduce this deviation, such that oscillations from the N\'eel state can be improved and interpreted as precession of a large ($s=N/2$) spin~\cite{Choi2018}. 

\begin{figure}[h]
    \centering
    \includegraphics[width=0.5\textwidth]{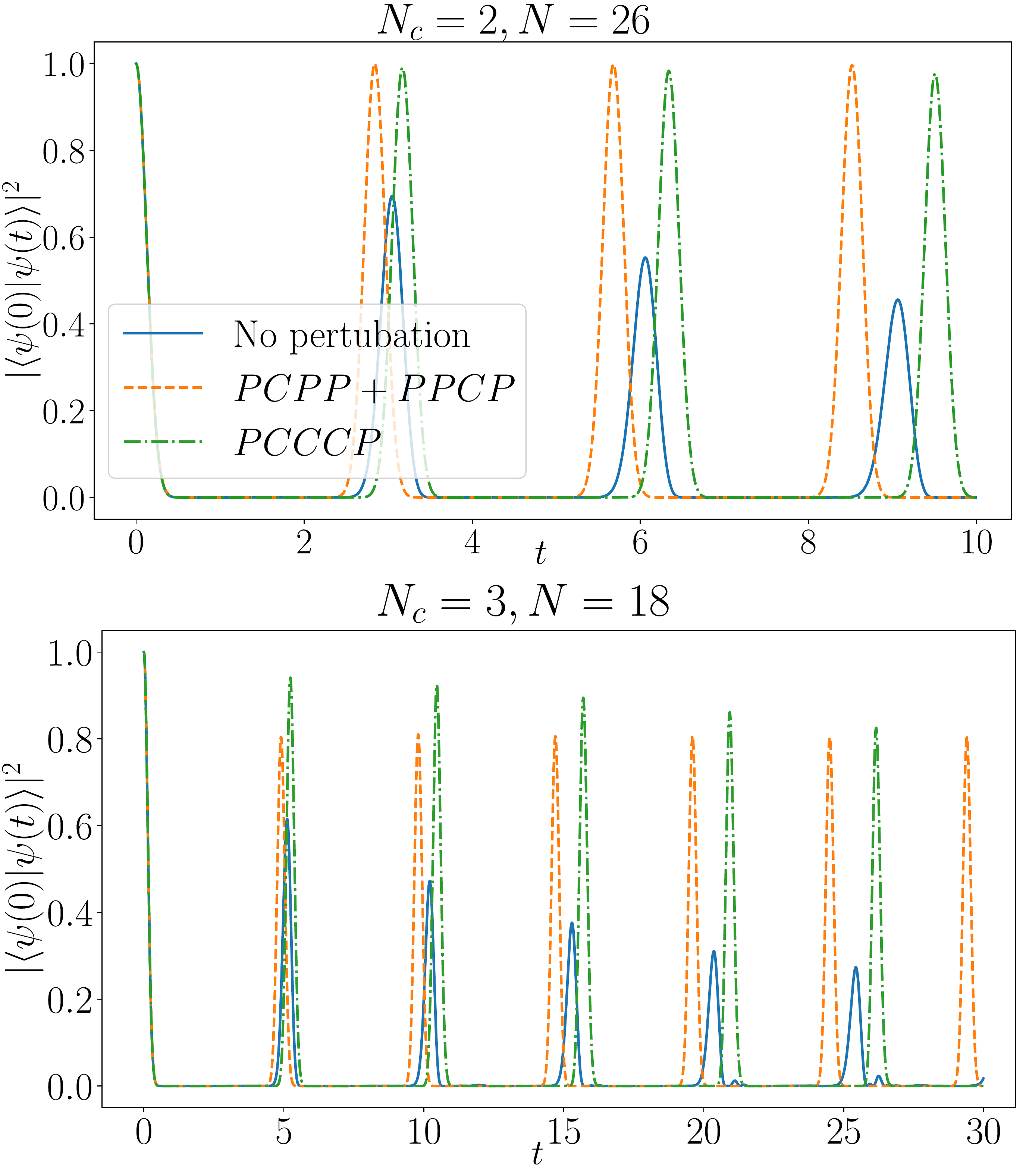}
    \caption{Effect of pertubations $H_1^{PCPP}$ in Eq.~(\ref{eq:pert1}) and  $H_3^{PCCCP}$ in Eq.~(\ref{eq:pert2}) on the revivals from a N\'eel like state. Results are for PXP spin-1/2 and $N_c=3$ colored clock models, with periodic boundary conditions.}
    \label{fig:}
\end{figure}

Further, we have numerically found a different family of perturbations
\begin{eqnarray}\label{eq:pert2}
 H_d^{PXXXP} = \sum_n P_{n-1} X_{n} X_{n+1},...,X_{n+d-1} P_{n+d},\;\;\;\;\;
\end{eqnarray}
 with $d=3,5,7,\ldots$, which also improves revivals.  Taking the same form of perturbations, but letting $X \rightarrow C$, revivals in clock models can also be substantially improved. We find the optimal coefficients for the 1st order perturbations using golden section search, taking as a cost function the first revival fidelity deficit, $1-F=1-\vert \langle \psi(0) \vert \psi(t) \rangle \vert^2$. Optimal values for $N_c=2,N=22$ were $0.10856$ (PPXP), $0.12296$ (PXXXP), while for $N_c=3,N=16$ optimal values were $0.09443$ (PPCP), $0.09112$ (PCCCP). 

We note here that PPCP perturbations appear to improve the first fidelity revival to some threshold, and revivals no longer decay after this. Intuitively, this suggest all components of the wave function not residing in some subspace is rapidly shed, followed by perfect oscillations in this subspace. Via a rotation, we know $H=PCP+\lambda (PPCP+PCPP)=P^{'} X P^{'} + \lambda (P^{'} P^{'} X P^{'} + P^{'} X P^{'} P^{'})$. 
Thus we also expect an emergent $SU(2)$ subspace, whose algebra is specified by $H^{+}$,$H^{-}$ now given in terms of $P^{'}$. However, in the rotated basis, the initial state $\vert 0101...\rangle_{clock}$ is now a highly entangled state. Thus it appears any component not in the emergent $SU(2)$ subspace will dephase away and we are left with perfect revivals in this subspace. 

In contrast, PCCCP perturbations can improve fidelity revivals to a greater degree for short times, but fidelity can still be observed to decay at longer times. This suggests the improvements comes from an orthogonal mechanism, not an emergent SU(2) symmetry. We leave this question for future work.

\subsection{Proximity to Integrability}

It has been suggested~\cite{Khemani2018} that $H_d^{PPXP}$ perturbations, Eq.~(\ref{eq:pert1}), drive the spin-1/2 PXP Hamiltonian to an integrable point. The argument for this was the numerically observed minimum in the mean level statistics parameter $\langle r \rangle$~\cite{OganesyanHuse}, tending towards the Poisson value $\approx 0.39$.  This can be observed in Fig~\ref{fig:pxp_level_stat_scan}, where we scan $\langle r \rangle$, resolved in the $k=0$, inversion symmetric sector, while varying the strength of perturbation. Vertical lines mark the optimal perturbation strength when optimizing for fidelity revivals. For PXP, both perturbations, PPXP and PXXXP, result in a dip of $\langle r \rangle \approx 0.39$, although the optimal perturbation strength with respect to fidelity moves the system further from this minimum~\cite{Choi2018}.  
\begin{figure}[h]
    \centering
    \includegraphics[width=0.45\textwidth]{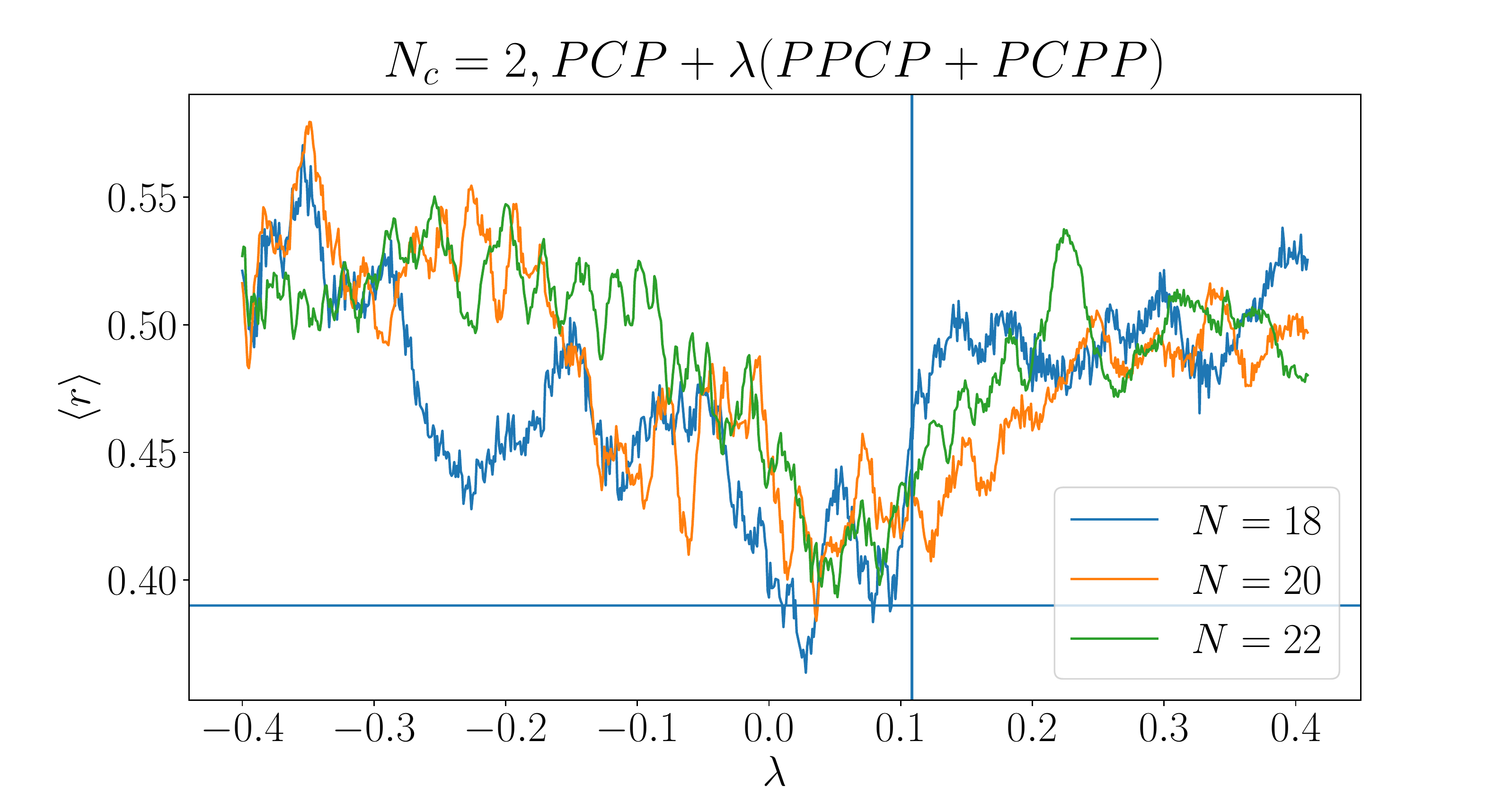}
    \includegraphics[width=0.45\textwidth]{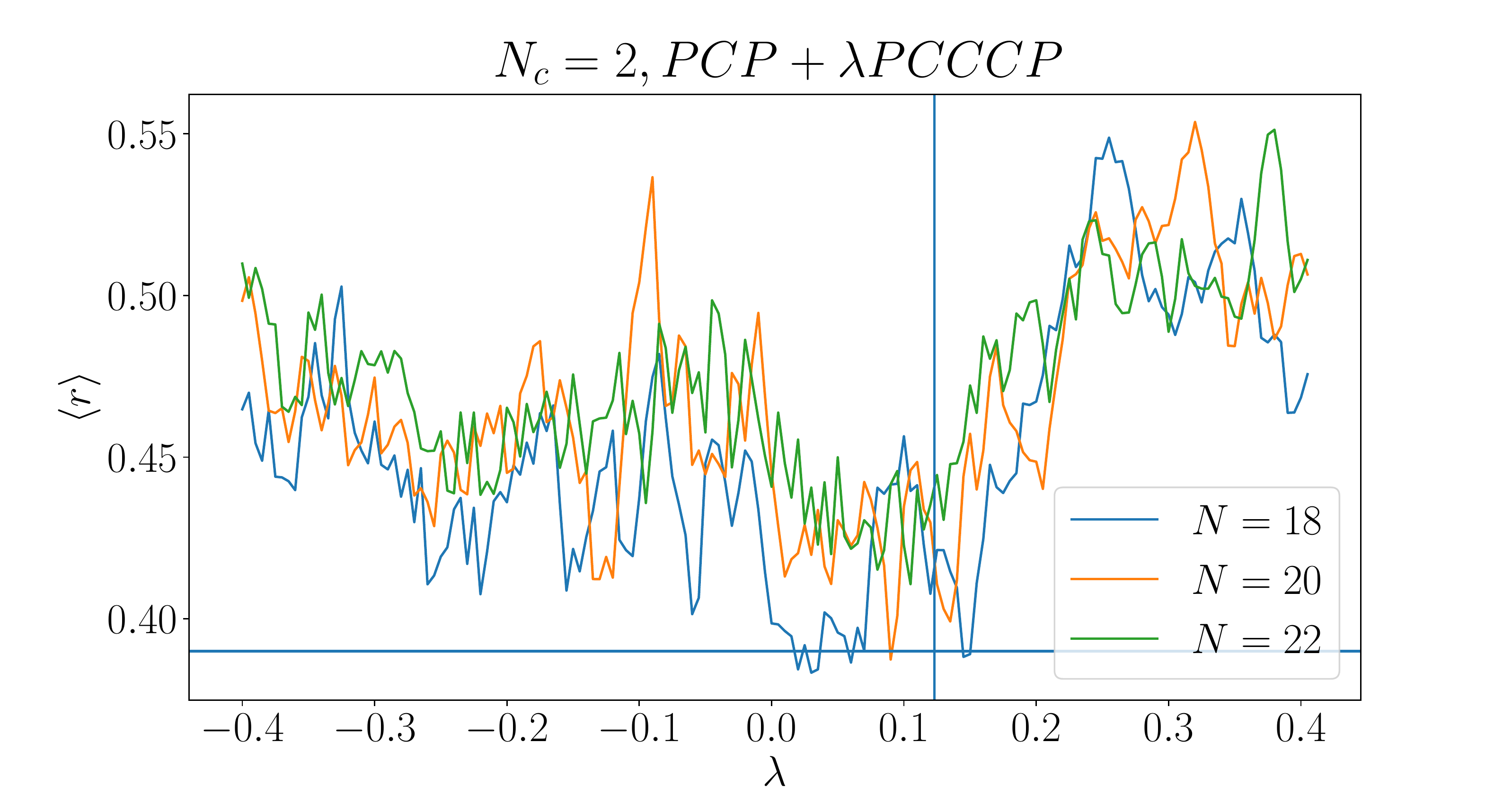}
    \caption{Effect of perturbation strength on the mean level statistics parameter $\langle r \rangle$ for the spin-1/2 PXP model. Top: $H_d^{PPXP}$,  bottom: $H_d^{PXXXP}$.}
    \label{fig:pxp_level_stat_scan}
\end{figure}
\begin{figure}[h]
    \centering
    \includegraphics[width=0.45\textwidth]{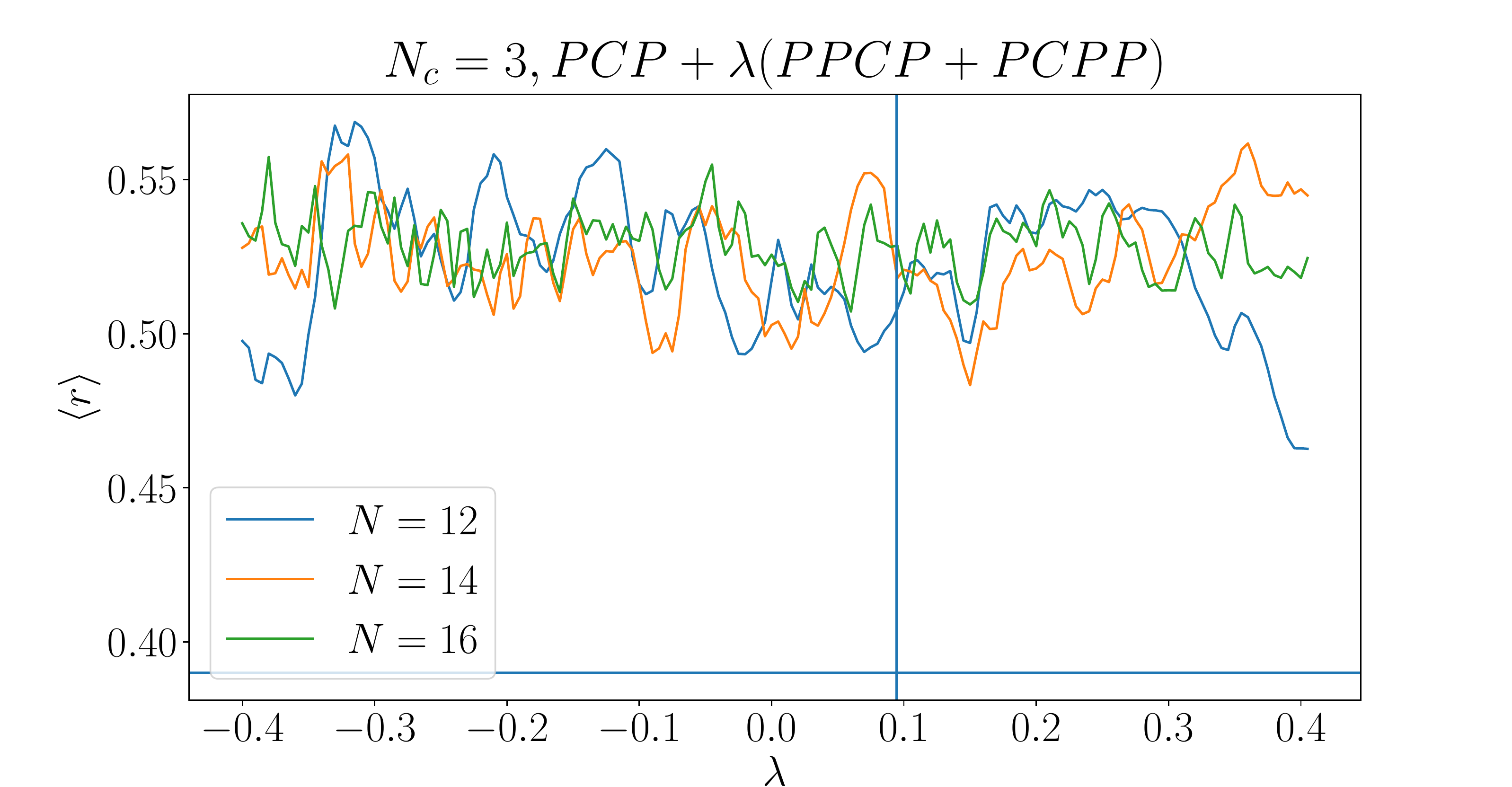}
    \includegraphics[width=0.45\textwidth]{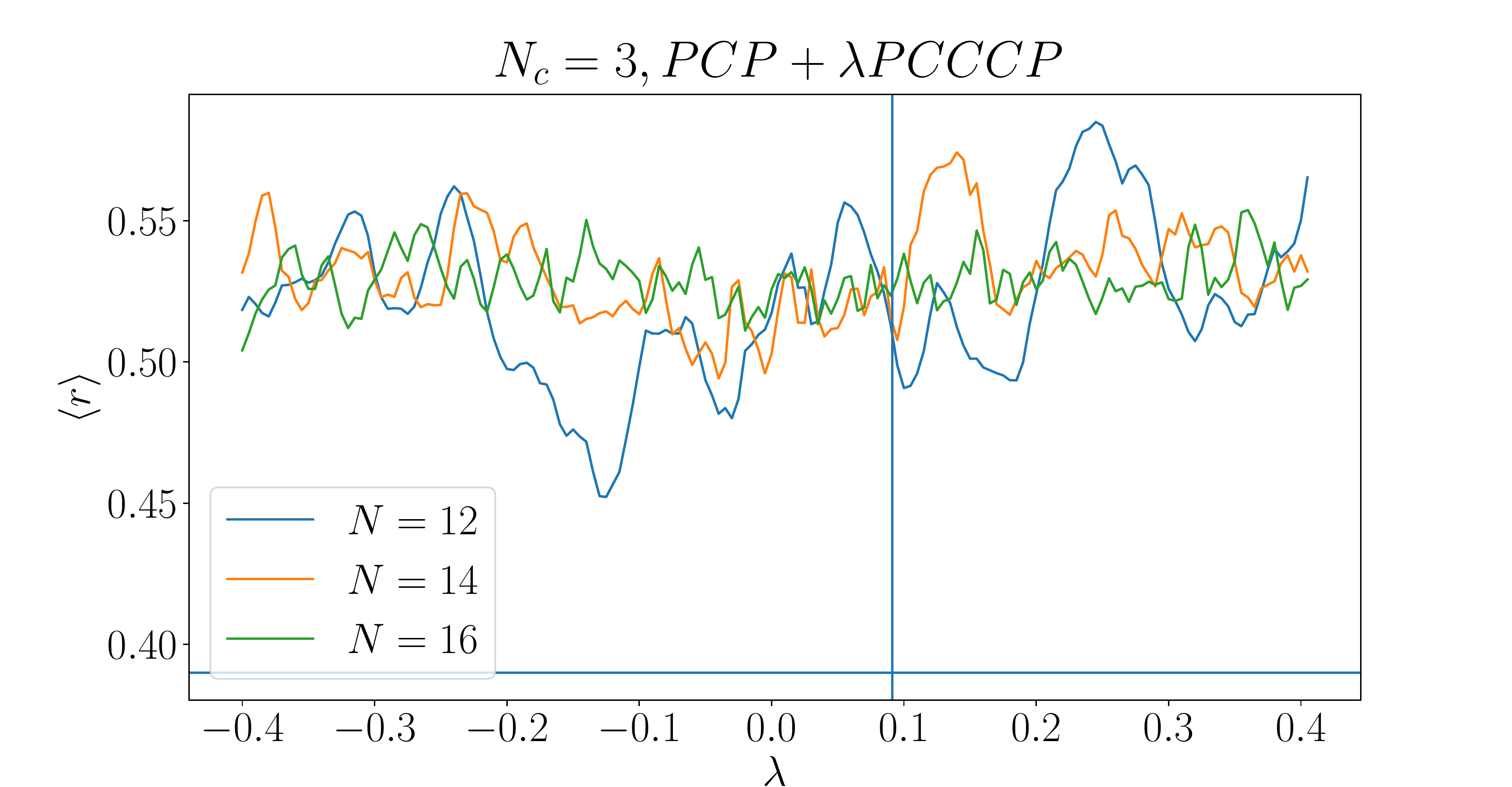}
    \caption{Effect of perturbation strength on the mean level statistics parameter $\langle r \rangle$ for the $N_c=3$ colored clock model. Top: $H_d^{PPXP}$, bottom: $H_d^{PXXXP}$.}
    \label{fig:pcp_level_stat_scan}
\end{figure}

In contrast, no similar features is observed for the $N_c=3$ colored clock (Fig~\ref{fig:pcp_level_stat_scan}), where $\langle r \rangle \approx 0.53$ for all values scanned, including no perturbation and the optimal perturbation. While we cannot rule out a proximate integrable point for the clock models after only scanning along two axes of the vector space of all possible perturbations, it seems that optimal revivals are not correlated with integrability.

\end{document}